\documentclass[aps,prd,amsmath,floats,floatfix,
twocolumn,
superscriptaddress,nofootinbib,showpacs]{revtex4-1}

\usepackage[colorlinks, pdfborder={0 0 0}, plainpages=false]{hyperref}
\usepackage{graphicx}
\usepackage{xspace}
\usepackage[usenames,dvipsnames]{color}
\usepackage[dvipsnames]{xcolor}
\usepackage{amssymb}
\usepackage{empheq}
\usepackage{placeins}
\usepackage[normalem]{ulem}
\usepackage[utf8]{inputenc}

\newcommand{\Olap}{\mathcal{O}}
\newcommand{\D}{\mathrm{d}}
\newcommand{\Eff}{\mathrm{FF}}
\newcommand*{\defeq}{\mathrel{\vcenter{\baselineskip0.5ex \lineskiplimit0pt
                     \hbox{\scriptsize.}\hbox{\scriptsize.}}}%
                     =}
\newcommand{\mchirp}{\mathcal{M}_c}
\newcommand{\chieff}{\chi_\mathrm{eff}}
\newcommand{\chimw}{\chi_\mathrm{mw}}

\newcommand{\AEI}{\affiliation{Max Planck Institute for Gravitational Physics (Albert Einstein Institute), Am M{\"u}hlenberg~1, 14476 Potsdam-Golm, Germany}}
\newcommand{\Caltech}{\affiliation{Theoretical Astrophysics 350-17,
    California Institute of Technology, Pasadena, CA 91125, USA}}
\newcommand{\Cornell}{\affiliation{Cornell Center for Astrophysics and Planetary Science, Cornell University, Ithaca, New York 14853, USA}}
\newcommand{\CITA}{\affiliation{Canadian Institute for Theoretical
    Astrophysics, 60 St.~George Street, University of Toronto,
    Toronto, ON M5S 3H8, Canada}} %
\newcommand{\TorontoPhysics}{\affiliation{Department of Physics, University of Toronto, 60 St.~George Street, Toronto, ON M5S 3H8, Canada}} %
\newcommand{\CIFAR}{\affiliation{Canadian Institute for Advanced Research, 
    180 Dundas St.~West, Toronto, ON M5G 1Z8, Canada}} %
\newcommand{\Princeton}{\affiliation{Department of Physics, 
	Princeton University, Jadwin Hall, Princeton, NJ 08544, USA}}
\newcommand{\JPL}{\affiliation{Jet Propulsion Laboratory, California Institute of Technology, 4800 Oak Grove Drive, Pasadena, CA 91109, USA}}

\begin{document}

\title{
Accuracy of binary black hole waveform models for aligned-spin binaries
}

\author{Prayush Kumar}\CITA
\author{Tony Chu}\Princeton
\author{Heather Fong}\CITA\TorontoPhysics
\author{Harald P. Pfeiffer}\CITA\AEI\CIFAR
\author{Michael Boyle}\Cornell
\author{Daniel A. Hemberger}\Caltech
\author{Lawrence E.~Kidder}\Cornell
\author{Mark A.~Scheel}\Caltech
\author{Bela~Szilagyi}\Caltech\JPL

\date{\today}

\begin{abstract}
  Coalescing binary black holes are among the primary science targets
  for second generation ground-based gravitational wave detectors.
  Reliable gravitational waveform models are central to detection of
  such systems and subsequent parameter estimation.  This paper
  performs a comprehensive analysis of the accuracy of four waveform
  models for binary black holes with aligned spins, utilizing a new
  set of $84$ high-accuracy numerical relativity simulations. 
  Our analysis covers comparable mass binaries (mass-ratio $1\le q\le 3$),
  and samples independently both black hole spins up to dimensionless
  spin-magnitude of $0.9$ for equal-mass binaries and $0.85$ for
  unequal mass binaries.  Furthermore, we focus on the high-mass
  regime (total mass $\gtrsim 50M_\odot$).
  We find that the PhenomD and SEOBNRv2 models perform very
  well for signal detection, losing less than $0.5\%$ of the
  recoverable signal-to-noise ratio, except that
  SEOBNRv2's
  efficiency drops slightly for both black holes spins aligned at
  large magnitude. For parameter estimation, PhenomD and SEOBNRv2 are 
  satisfactory for moderately strong signals, although accuracy deteriorates with
  increased mass-ratio, and when at least one black hole spin is large
  and aligned.
  PhenomD agrees generally even better with the NR simulations than SEOBNRv2, 
  with the latter deviating with the NR merger phase at $q=2,3$ and 
  the highest aligned spins. The PhenomC and SEOBNRv1 models are
  found to be distinctly less accurate than their more recent 
  counterparts.  
  Finally, we quantify the systematic bias expected from all
  four waveform models during parameter estimation for several
  recovered binary parameters: chirp mass, mass-ratio, and effective spin.
\end{abstract}


\maketitle

\section{Introduction}

Gravitational-wave (GW) astronomy enters an exiting time 
with a global concerted effort going online to detect gravitational waves
with ground-based facilities.  In North America, the
Advanced Laser Interferometer Gravitational-wave Observatory
(aLIGO) operates two $4-$km scale GW detectors\cite{Harry:2010zz,aLIGO2},
located in Hanford, Washington and Livingston, Louisiana. Both of
these instruments began their first observation run ``O1'' in
September 2015, which is scheduled to last for four 
months~\cite{Aasi:2013wya}, operating at more than three times the strain 
sensitivity of the initial LIGO detectors~\cite{Aasi:2014mqd}.
In addition, the upgrades to the Virgo
detector~\cite{TheVirgo:2014hva}, construction of the KAGRA
detector~\cite{Somiya:2012,kagra}, and planning of LIGO-India
detector~\cite{2013IJMPD..2241010U} are underway.

Binary black holes (BBHs) are among the most promising GW sources
for detection with aLIGO. Compact binary merger rate estimates suggest
a GW detection rate of approximately a few tens of binary black holes
(BBH) every year~\cite{Abadie:2010cfa}. The actual masses of
astrophysical black holes are uncertain, but observations and
population synthesis studies suggest that BHs formed from stellar
core-collapse can have masses up to and higher than
$34M_\odot$~\cite{2008ApJ...678L..17S,2010ApJ...714.1217B}.  Also,
recent measurements using continuum fitting and X-ray reflection
fitting suggest that black holes can have high spin, with the BH
angular momentum in dimensionless units exceeding
$0.8$~\cite{2011ApJ...742...85G,2012MNRAS.424..217F,Gou:2013dna,
  2009ApJ...697..900M,McClintock:2006xd,Miller:2009cw,McClintock:2013vwa}.
Therefore the observations of GWs emitted by spinning BBHs will allow us to
understand the spin-spin and spin-orbit dynamics of the two-body
system, apart from allowing us to test strong-field dynamics of
General Relativity.
Unlocking the full scientific potential of BBH GW observations, however,
will require us to detect as many such GWs as possible, and to
accurately characterize and classify the BBH systems that emitted them.

Optimal GW searches for stellar-mass BBH signals are based on
matched-filtering the detector data with modeled waveforms. Past
LIGO-Virgo searches for compact binaries used models of non-spinning
BBH inspirals as filtering templates,
e.g.~\cite{Colaboration:2011nz,Abadie:2010yb,
  Abbott:2009qj,Abbott:2009tt} (with the exception
of~\cite{Abbott:2007ai}).  Recent progress has moved the collaboration
towards using inspiral-merger-ringdown models of aligned-spin BBHs as
filters. It has been shown that doing so will significantly increase
search efficiency against generically oriented
binaries~\cite{Harry:2013tca}.
Furthermore, it has been shown that complete inspiral-merger-ringdown
(IMR) waveforms are needed for the observation of BBHs with
$M\gtrsim 12M_\odot$~\cite{Brown:2012nn}.  It is therefore important
for the aligned-spin candidate waveform models to be carefully
examined for accuracy in capturing the entire coalescence process,
including merger-ringdown.
Early work on assessing the accuracy of different waveform models has
focused on model-model comparisons~\cite{Damour:2000zb,
  Damour2001,Damour:2002kr,Damour02,Gopakumar:2007vh,Buonanno:2009}. In
absence of more accurate reference waveforms, such studies have been
limited by the most accurate model they consider, and have used
model-model agreement to make statements about model accuracy.
More recently, there have been extensive studies of waveform models
involving high-accuracy numerical relativity (NR)
simulations~\cite{Boyle2007,
  Boyle2008b,Boyle:2008,Pan2007,Hannam2007c,Hannam:2010ec,HannamEtAl:2010,
  MacDonald:2012mp,Hinder:2013oqa,Kumar:2015tha}.  However, most of
these investigations have focused on binaries with zero spins or
modest spin magnitudes. Furthermore, while recently published
models~\cite{Taracchini:2013rva,Khan:2015jqa,Damour:2014sva} have used an 
unprecedented amount of information from NR to increase the accuracy of
their merger description~\cite{Husa:2015,Mroue:2013PRL,Hinder:2013oqa,Damour:2013tla},
their accuracy has not been investigated in a systematic manner over the BBH 
parameter space.

\begin{figure*}
\centering
\includegraphics[width=\textwidth,trim=15 17 15 17]{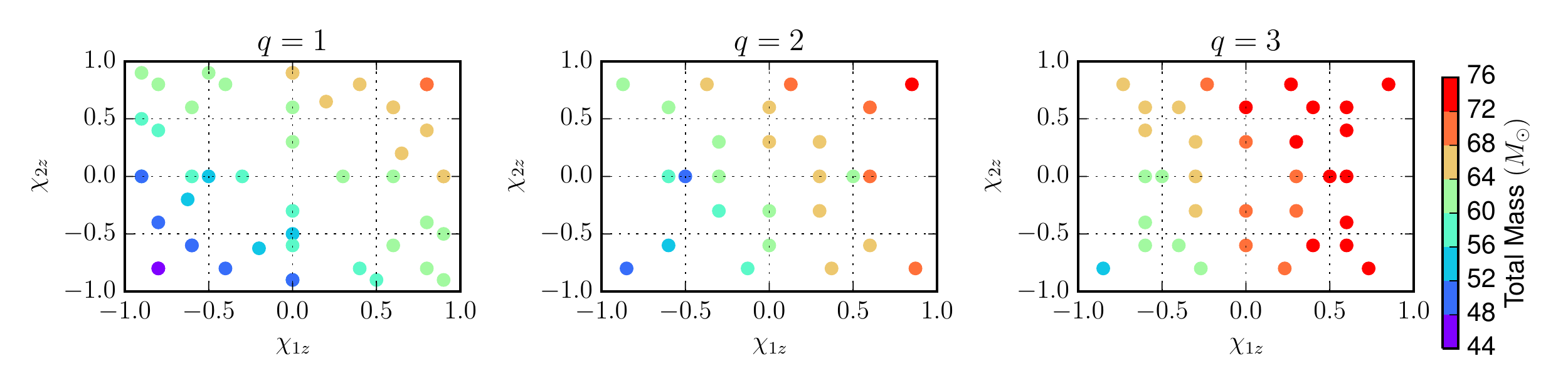}
\caption{Parameter space coverage of the simulations considered here. For
mass-ratio $q=\{1,2,3\}$ we indicate the spin-components 
$(\chi_{1}, \chi_{2})$
projected onto the orbital angular momentum. Each point is color-coded by the 
lowest total mass to which the waveform can be scaled, such that the initial GW
frequency remains $\gtrsim 15$~Hz. In the $q=1$ panel, each simulation is 
plotted twice at $(\chi_{1}, \chi_{2})\rightarrow (\chi_{2}, \chi_{1})$ to
represent the symmetry under exchange of the two objects.
}
\label{fig:Min_TotalMass_Spin1z_Spin2z}
\end{figure*}

In this paper, we explore the accuracy of currently available
BBH waveform models using new high-accuracy NR simulations, 
from the perspective of their application to GW astronomy. 
The $84$ numerical waveforms were computed with the Spectral Einstein Code
(SpEC)~\cite{SpECwebsite} and are presented in detail in a companion 
paper~\cite{Chu:2015kft}.
This catalog covers non-precessing configurations, i.e. BBHs
with spin-vectors parallel or anti-parallel to the orbital 
angular momentum.
More specifically, it spans mass-ratios $q\equiv m_1/m_2\in[1,3]$, and and spin-projection $\chi_i\equiv \vec{\chi}_i\cdot\hat L\in [-0.9,+0.9]$, where $i=1,2$ labels the two black holes, with mass $m_i$ and dimensionless angular momentum
$\vec\chi_i\equiv \vec S_i/m_i^2$, and where $\hat L$ denotes the
unit vector along the direction of the orbital angular momentum.
The median length of these simulations is 24 orbits, allowing us to
extend our comparisons down to binary masses as low as $40-70M_\odot$
(depending on configuration,
c.f. Fig.~\ref{fig:Min_TotalMass_Spin1z_Spin2z}) while still covering
aLIGO's frequency band above $15$~Hz.  We restrict probed total masses
below $150M_\odot$.

The waveform models we investigate include two NR-calibrated
Effective-One-Body (EOB) models (namely, SEOBNRv1 and
SEOBNRv2)~\cite{Taracchini:2012,Taracchini:2013rva}, and two recent
phenomenological models (namely IMRPhenomC and
IMRPhenomD)~\cite{Santamaria:2010yb,Khan:2015jqa}.
Both EOB and IMRPhenom models are constructed using (different)
extensions of Post-Newtonian (PN) dynamics of compact binaries, with free parameters
that are calibrated to NR simulations. We probe their accuracy in
different corners of the component spin space in this work.
The four waveform models were published over the period from 2010 to 2015. 
Given the rapid progress in waveform modeling and numerical relativity, we
expect newer models to be superior to older ones.
We model detector sensitivity using the zero-detuning high-power noise
power spectral density for aLIGO~\cite{Shoemaker2009}, and use
$f_\mathrm{low}=15$~Hz as the lower frequency cutoff for filtering.

We perform the following studies. First, we measure the faithfulness
of different waveform models by calculating their noise-weighted
overlaps against the new NR waveforms.  We find that (i) both SEOBNRv2
and IMRPhenomD are faithful to our NR simulations over most of the
spin and mass-ratio parameter space (overlaps $>99\%$), with overlaps
falling to $97-98\%$ when component spins are anti-parallel to each
other. However, when both BHs have large positive-aligned spins,
IMRPhenomD fares significantly better, while the overlaps
between SEOBNRv2 and NR fall to $80\%$; (ii) both SEOBNRv1 and
IMRPhenomC show larger disagreement with NR, and we clearly show that
they have been superseded by their respective recent versions in
accuracy. Specifically, we find that SEOBNRv1 deteriorates when the
spin on the \textit{larger} BH is $\gtrsim+0.5$ (with overlaps falling
to $80\%$), while IMRPhenomC performs poorly when the spin magnitude
on the \textit{smaller} BH exceeds $\approx 0.5$, with overlaps
falling below $80\%$.  While we do not find a strong correlation
between model accuracy and mass-ratio for the SEOBNRv2 and
IMRPhenomD models, we do find that both SEOBNRv1 and IMRPhenomC
deteriorate in accuracy with increasing binary mass-ratio.
In addition, we also show that the IMRPhenomD and SEOBNRv2 models are
indistinguishable from NR simulations in large regions of the considered 
parameter space up an effective signal-to-noise
(SNR) of $20$ and $15$, respectively, albeit with significant dependence
on the mass-ratio and spins.

In our second study, we assess the viability of waveform
waveform models for aLIGO \textit{detection} searches
for high-mass BBHs.  We compute the overlaps between each
rescaled NR waveform and a large set of model waveforms that sample
the binary mass and spin parameter space densely. From this, we
recover the maximum fraction of the optimal signal SNR that any
waveform model can recover -- with the only loss being caused by
intrinsic inaccuracies of the model itself.
We find that (i) both SEOBNRv2 and IMRPhenomD recover more than
$99.5\%$ of the optimal SNR over most of the mass and spin parameter
space, except when both BHs have large aligned spins, where the
inaccuracies of SEOBNRv2 lead to a drop in SNR recovery to $97\%$ of
its optimal value; and (ii) both IMRPhenomC and SEOBNRv1 compensate
for their intrinsic inaccuracy with maximization of SNR over waveform
parameters, recovering $>98\%$ of the optimal SNR over most of the
parameter space considered within their domain of applicability. This
is a manifestation of the efficient utilization of the intrinsic mass
and spin degeneracy of gravitational waveforms~\cite{Baird:2012cu,
Purrer:2015nkh}, allowing IMRPhenomC to
be a fairly \textit{effectual} model despite being unable to reproduce
NR waveforms with identical masses and spins.
Overall, we conclude that both SEOBNRv2 and IMRPhenomD are viable
for modeling waveforms in aLIGO searches aimed at comparable
mass-ratio high-mass BBHs. This validates the use of SEOBNRv2 by
current and future aLIGO searches.
We note that due to high computational cost of evaluating the 
SEOBNRv2 model, aLIGO data analyses use its reduced-order
model~\cite{Purrer:2015tud} which mitigates this drawback.

Our third study concerns BBH parameter estimation from GW signals,
which, when accurately done, will provide unique insight into
astrophysical processes involving stellar-evolution, compact binary
formation and evolution~\cite{Read2009b,
  Kreidberg:2012,Fairhurst:2009tc,2011ApJ...739...99N,
  2012A&A...539A.124L,2012A&A...541A.155A,2012ApJS..203...28E,
  2013ApJ...767..124N,metzger:11,2012IAUS..285..358M,Li:2011cg,Li:2011vx,
  Littenberg:2015tpa,Mandel:2015spa}. Full Bayesian analyses of GW
signals require models that faithfully reproduce real GWs in order to
map them back to the properties of their source binaries. Model
inaccuracies manifest themselves as biases in the recovered values of
the mass and spin parameters of BBHs. Therefore, we investigate the
level of systematic biases that using different (aforementioned)
inspiral-merger-ringdown (IMR) waveform models will incur.
We find that: 
(i) binary chirp-mass is best recovered by IMRPhenomD
(within $\pm 2-5\%$), especially for spin-aligned systems. For systems
with anti-aligned spins, the systematic bias in chirp-mass is similar
for both IMRPhenomD and SEOBNRv2, rising above $5\%$ at the higher end
of the sampled binary mass range. 
(ii) total mass is recovered with similar accuracy ($2-5\%$) by both
SEOBNRv2 and IMRPhenomD, while not as well as both recover $\mchirp$. The
older SEOBNRv1 and IMRPhenomC models, while furnishing larger biases overall,
recover $M$ better than $\mchirp$.
(iii) Binary mass-ratio is also best
recovered by IMRPhenomD (within $10-15\%$), with SEOBNRv2
systematically \textit{under}-estimating mass-ratios for binaries with
anti-aligned spins, and \textit{over}-estimating for positive-aligned
spins (by up to $\pm 20\%$).
(iv) We test the recovery of the PN effective-spin combination
$\chi_\mathrm{eff}$ that appears at leading-order in inspiral phasing.
As with the mass parameters, we find that IMRPhenomD recovers
$\chi_\mathrm{eff}$ best (within $\pm 0.1$), especially for strongly
spin-aligned binaries. While SEOBNRv2 shows marginally higher spin
biases (up to $\pm 0.15$) for
high-mass binaries with $M\gtrsim 100M_\odot$, both SEOBNRv1 and
IMRPhenomC models incur higher biases in spin recovery (up to
$\pm 0.25$) over different regions of the parameter space.
Overall, we find that both SEOBNRv2 and IMRPhenomD have comparable
accuracy in terms of parameter recovery, with IMRPhenomD performing
better of the two for binaries with large aligned $\chi_\mathrm{eff}$
and/or high masses.

We note that a recent study~\cite{Purrer:2015nkh} shows that the biases we
find for SEOBNRv2 will become comparable to statistical uncertainty in 
spin recovery only at SNRs $\approx 20-30$.
However, a more detailed MCMC analysis will be needed to (i) determine
the same for highly spinning binaries, where SEOBNRv2 deviates 
significantly from NR, and (ii) to understand the impact of the systematic
biases for the IMRPhenomD model that we report here, by comparing them
with the associated statistical uncertainty in parameter recovery.
We also recall that the present study applies to high-mass BBHs, with
total-masses $\gtrsim 50M_\odot$.  At lower binary masses, the NR
waveforms do no longer cover the entire aLIGO frequency band, and one
needs either longer NR simulations or one needs to hybridize the
existing simulations with PN inspiral waveforms. We also note that we
plan to follow-up the interesting patterns seen in the
high-spin/high-spin corner of the BBH parameter space in the future in
order to better understand the accuracy of analytical models there.

The remainder of the paper is organized as follows.
Sec.~\ref{s1:Methodology} summarizes the salient features of the new
catalog of NR simulations used in this analysis, describes different
measures of waveform-model accuracy, and summarizes the different
waveform models analyzed in this paper. In Sec.~\ref{s1:faithfulness}
we present overlap comparisons of different waveform models with our
NR waveforms. In Sec.~\ref{s1:effectualness} we measure the efficacy
of different waveform models as detection filters. In
Sec.~\ref{s1:param_bias} we analyze the systematic biases in the
recovery of binary mass and spin parameters, associated with the
different waveform models we consider in this paper.  Finally, in
Sec.~\ref{s1:conclusions} we summarize and discuss our results.

\section{Methodology}\label{s1:Methodology}

\subsection{Numerical Relativity Simulations}\label{s2:numrel_simulations}
The BBH simulations considered here were performed with the Spectral Einstein 
Code (\texttt{SpEC})~\cite{SpECwebsite}, and were presented 
in~\cite{Chu:2015kft}. Initial data is constructed with the pseudo-spectral 
elliptic solver described in~\cite{Pfeiffer2003}, using the extended conformal 
thin-sandwich method~\cite{Yo2004} with quasi-equilibrium boundary 
conditions~\cite{Cook2004}.
Evolutions use a first-order representation of the generalized harmonic 
system~\cite{Friedrich1985,Garfinkle2002,Pretorius2005c,Lindblom:2007} with 
a damped-harmonic gauge~\cite{Szilagyi:2009qz}. The computational grid is 
adaptively refined~\cite{Szilagyi:2014fna}, and the excision boundaries 
are dynamically adjusted to follow 
the apparent horizons~\cite{Scheel2009,Szilagyi:2009qz,Hemberger:2012jz}. 
Interdomain boundary conditions are enforced with 
a penalty method~\cite{Gottlieb2001,Hesthaven2000}, 
and constraint-preserving outgoing-wave 
conditions~\cite{Lindblom2006,Rinne2006,Rinne2007} are imposed at the 
outer boundary.

Our simulations consist of 84 configurations at mass-ratios 
$q=m_1/m_2=\{1,2,3\}$. All simulations are non-precessing, i.e.
the dimensionless spin $\vec\chi_{1,2}$ of each hole is either aligned or 
anti-aligned with the direction of the orbital angular momentum $\hat{L}$.
The parameters of all simulations are plotted in 
Fig.~\ref{fig:Min_TotalMass_Spin1z_Spin2z}.
 22 simulations have only one hole spinning, 32 have both 
holes spinning with equal spin-magnitudes, and the remaining 30 have 
both holes spinning with unequal spin-magnitudes.
The spin components along $\hat{L}$ range over
$-0.9 \leq \chi_{1,2} \leq 0.9$. All evolutions have initial orbital 
eccentricity $e<10^{-4}$. The evolutions include an average of 24 orbits, 
with the shortest having 21.5 orbits and the longest having 32 orbits.
BBH waveforms can be rescaled to any total mass $M=m_1+m_2$.
Fig.~\ref{fig:Min_TotalMass_Spin1z_Spin2z} also indicates the lowest 
total mass $M_\mathrm{low}$ for each configuration, such that the rescaled
waveform covers the aLIGO frequency range for $f\geq f_\mathrm{low}=15$~Hz.

\subsection{Accuracy measures}\label{s2:accuracy_measures}

We can define an inner product between two waveforms $h_1$ and $h_2$ as
\begin{equation}\label{eq:innerproduct}
\langle h_1, h_2\rangle\equiv \int_{f_\mathrm{low}}^{f_\mathrm{high}} \dfrac{\tilde{h_1}(f)\,\tilde{h^*_2}(f)}{\mathrm{S}_n(|f|)}\D f,
\end{equation}
where $\tilde{h}(f)$ represents the Fourier transform of $h$, ${}^*$ in  
superscript represents complex conjugation, $\mathrm{S}_n(|f|)$ is the power 
spectral density of detector noise. We integrate the inner product over the 
frequency interval $[f_\mathrm{low}, f_\mathrm{high}]$, which spans the 
sensitive band of GW detector. In this paper we use $f_\mathrm{low}=15$~Hz,
$f_\mathrm{high}=4096$~Hz, and the zero-detuning high-power noise 
curve~\cite{Shoemaker2009} to model aLIGO at design sensitivity.
This inner product is sensitive to an arbitrary
phase and time shift between the two waveforms. Since both of these are extrinsic
parameters and of little astrophysical interest, we maximize the inner product 
over them to define the maximized overlap $\Olap$, 
\begin{equation}\label{eq:overlap}
\Olap\left(h_1,h_2\right) = \underset{\phi_0, t_0}{\mathrm{max}}\dfrac{\langle h_1, h_2\rangle}{\sqrt{\langle h_1, h_1\rangle\,\langle h_2, h_2\rangle}}.
\end{equation}
This overlap measures the correlation between any two given
waveforms. We use the overlap to measure the accuracy of analytical
waveform families by comparing to NR waveforms with identical physical
parameters. This assumes that the latter closely reproduce
\textit{true} waveforms in nature. The error analysis
in~\cite{Chu:2015kft} shows that numerical errors of the NR
waveforms cause mismatches $1-{\cal O}<5\times 10^{-4}$, with a median
value of $1-{\cal O}\sim 3\times 10^{-4}$.  Therefore, we expect that
overlaps computed here to be influenced by NR errors only for
$\mathcal{O}>0.9995$.

GW detection searches use a discrete set of waveforms, called a 
``template bank'', to filter detector data with. This bank spans the range of
mass and spin parameters considered in the search, and can be visualized as a 
multi-dimensional lattice. There are two sources of SNR loss from using 
template banks. First, the density of templates in the parameter space. This
is a free parameter which trades loss of SNR with the number of templates to
be searched with. Customarily, a $3\%$ loss in SNR is viewed as acceptable.
The second source of error -- the focus of this paper -- is the accuracy of
the underlying analytical waveform family that is used to generate the 
templates. The second source is somewhat compensated for by the freedom of 
maximizing the recovered SNR over intrinsic binary parameters, i.e., it 
does not matter \textit{which} template waveform fits a given signal in a detection
search. To investigate the SNR loss due to the
second factor alone, we compute the \textit{fitting factors} of different 
waveform models as follows.
For each combination $p$ of $(M=m_1+m_2, q=m_1/m_2, \chi_1, \chi_2)$ that we
rescale our NR waveforms to, we sample a set $\mathcal{S}_p$ of 
$8,000,000$
points in the vicinity of the true parameters ($p$) and compute the overlaps
between the NR waveform $h^\mathrm{NR}(p)$ and $M$odel waveforms 
$h^\mathrm{M}(i)$ for all points $i\in\mathcal{S}_p$. Finally, the 
fitting factor $\Eff$ of model $M$ for signal parameters $p$ is given by
\begin{equation}\label{eq:fittingfactor}
\Eff^\mathrm{M}(p)=\underset{i\,\in\,\mathcal{S}_p}{\mathrm{max}}\,\Olap\left(h^\mathrm{NR}(p), h^\mathrm{M}(i)\right).
\end{equation}
$\Eff$ is therefore the maximum fraction of the optimal SNR that a waveform model
can recover for a GW signal with parameters $p$. The deviation of fitting factor
from unity quantifies loss in SNR due model inaccuracy alone, and is in 
addition to any loss incurred due to the discreteness of the actual template 
bank used in a GW search.

\subsection{Waveform Models}\label{s2:waveform_models}

\begin{figure*}
\includegraphics[width=.5\columnwidth]{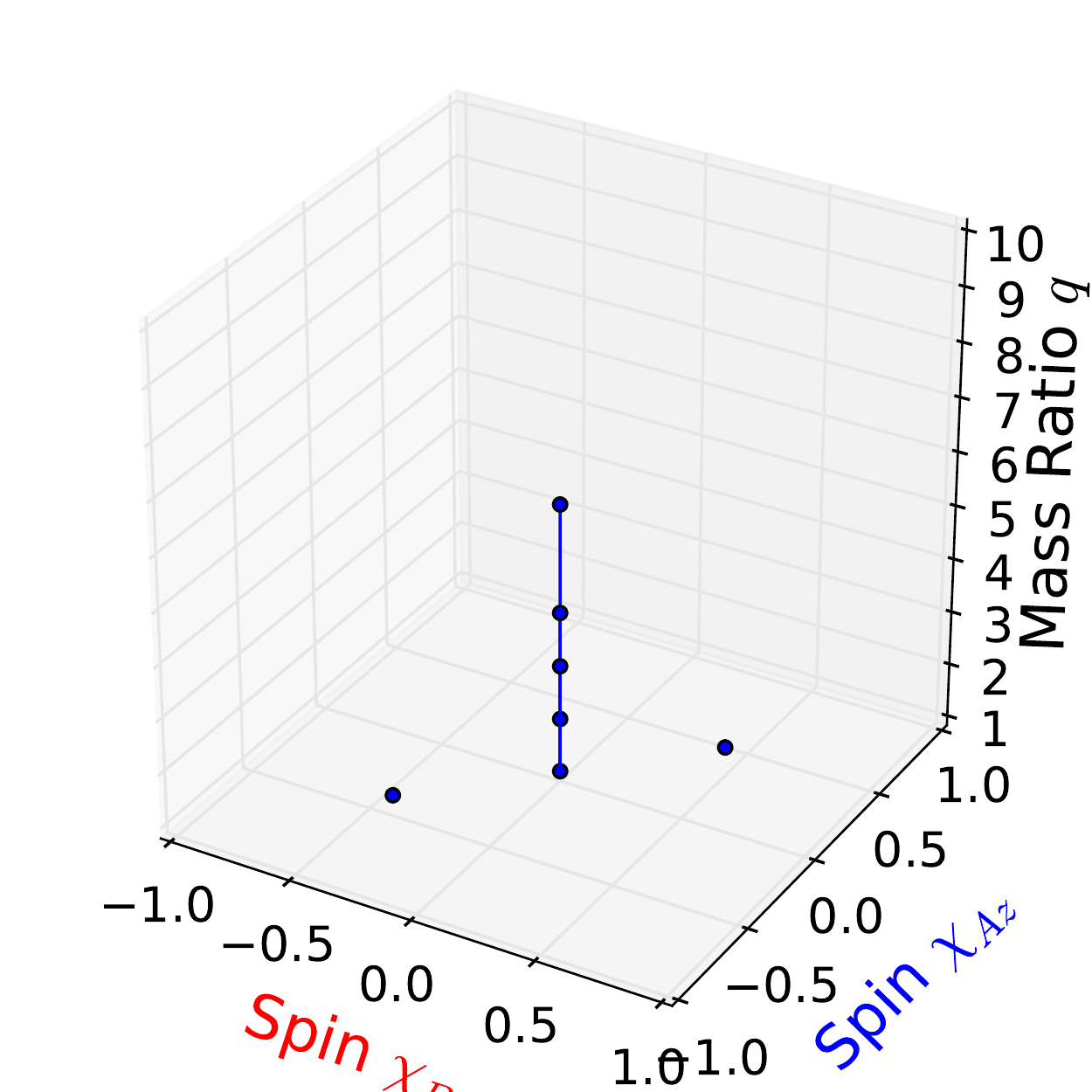}
\includegraphics[width=.5\columnwidth]{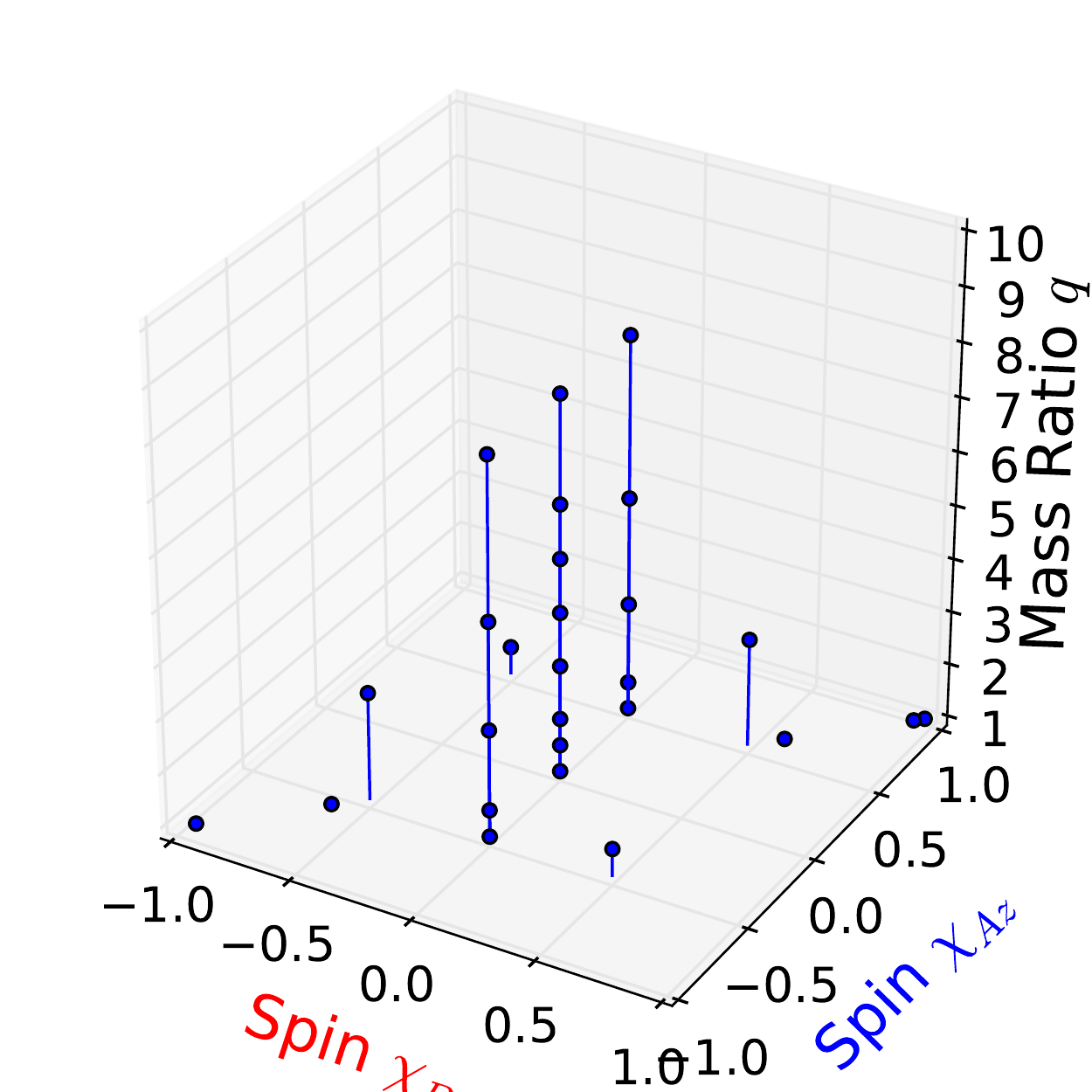}
\includegraphics[width=.5\columnwidth]{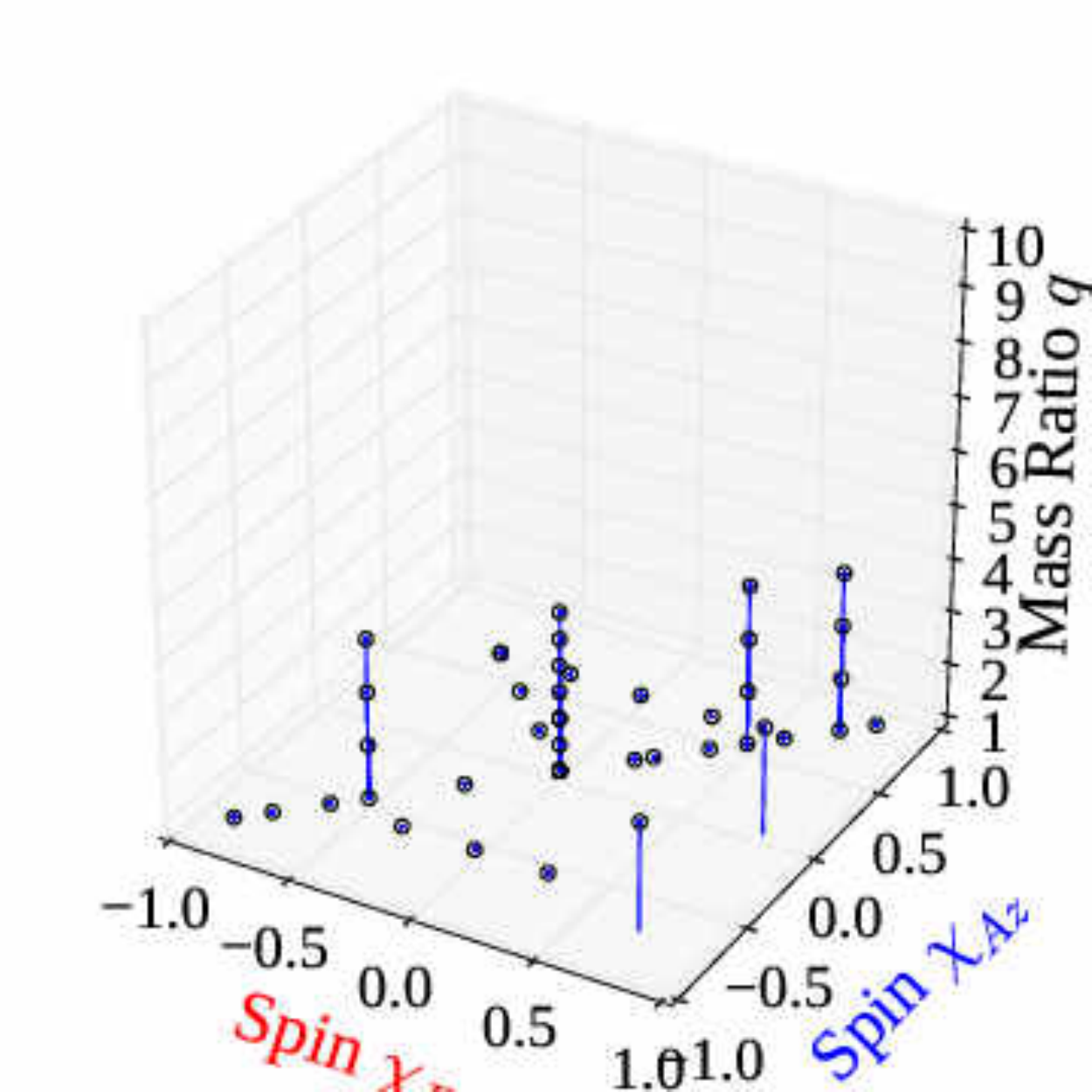}
\includegraphics[width=.5\columnwidth]{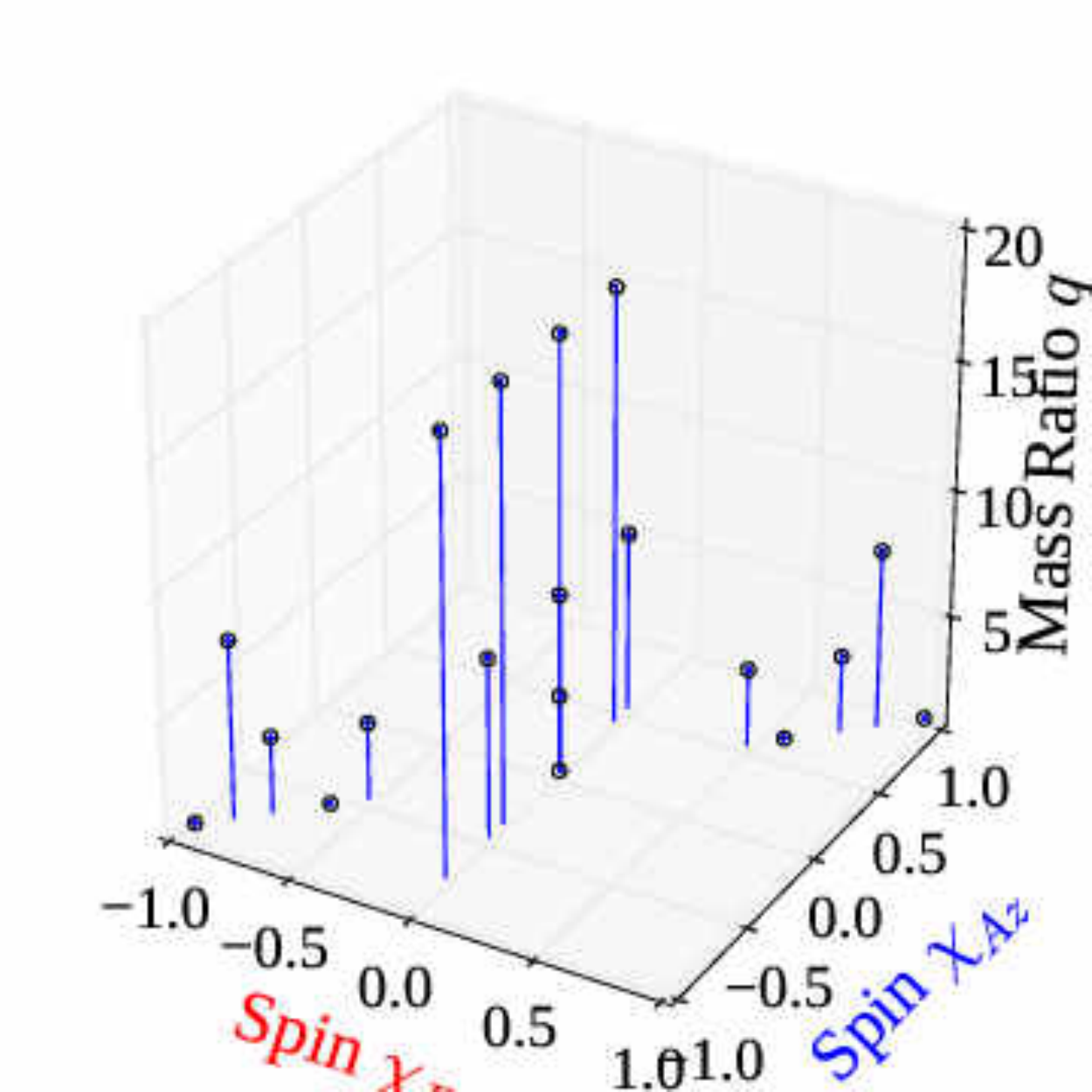}\\
\caption{Parameters of numerical-relativity simulations used to calibrate
the various inspiral-merger-ringdown models that we investigate in this 
paper, i.e. (left to right) SEOBNRv1, SEOBNRv2, IMRPhenomC and IMRPhenomD.
\label{fig:NRParams_IMR}
}
\end{figure*}

In this paper we investigate the following waveform models for aligned-spin
binary black holes.

\subsubsection{Effective-One-Body}
Buonanno and Damour~\cite{Buonanno99} developed an effective-one-body (EOB)
approach to the two-body problem in general relativity. Over the past decade
parameterized EOB models capable of describing the complete binary coalescence
process have been developed and calibrated using information 
from NR simulations~\cite{Buonanno99,Damour03,Damour2007,Damour2007a,DN2007b,
Damour:2007cb,Buonanno2007,DN2008,Buonanno:2009qa,Pan:2009wj,Buonanno:2009qa,
Barausse:2011dq,PanEtAl:2011,Taracchini:2012,Taracchini:2013rva,Damour:2013tla,
Damour:2014sva,Damour:2014sva}.
In the spin EOB framework, the dynamics of two compact objects of masses $m_1$
and $m_2$ and spins $\vec{\chi}_1$ and $\vec{\chi}_2$ is mapped onto the
dynamics of an effective particle of mass $\mu=m_1m_2/(m_1+m_2)$ and spin
$\vec{\chi}_*$ moving in a deformed-Kerr background with mass $M=m_1+m_2$
and spin $\vec{\chi}_\mathrm{Kerr}$. The parameterized spin mapping 
$\{\vec{\chi}_1,\vec{\chi}_2\}\rightarrow \vec{\chi}_*$ and the deformation of 
the background from Kerr is chosen to ensure that the inspiral dynamics of the 
test particle reproduce the PN-expanded dynamics of the original two-body
system. Free parameters are introduced into the models that represent unknown,
higher-order PN terms, or additional physical effects like corrections due to
non-circularity. Such free parameters are calibrated with NR simulations.
With the EOB system specified, a Hamiltonian $H_\mathrm{EOB}$
to describe its conservative dynamics can be written~\cite{Taracchini:2012,
Taracchini:2013rva}. The non-conservative dynamics is contained in a 
parameterized radiation-reaction term that is inserted in the equations of 
motion. This term sums over the outgoing GW modes and is
calibrated to reproduce NR simulations. The combination of these two 
pieces describe the binary inspiral through to merger, at which point 
a ringdown waveform is stitched on to the inspiral-merger 
waveform. This BH ringdown waveform is constructed as a linear superposition 
of the dominant quasi-normal modes (QNMs) of the Kerr BH formed at
merger~\cite{Berti2009,Buonanno:2009qa}, with amplitude and phase of each
QNM mode determined by the stitching process.

In this paper we focus on two aligned-spin EOB models which are 
calibrated to NR and used in contemporary LIGO data analyses: SEOBNRv1 and 
SEOBNRv2~\citep{Taracchini:2012,Taracchini:2013rva}. The SEOBNRv1 model has
been calibrated to five non-spinning simulations with $q=m_1/m_2=\{1,2,3,4,6\}$
and two equal-mass equal-spin simulations~\cite{Taracchini:2012}. It models binaries
with non-precessing BH spins in the range $-1\leq\chi_{1,2}\leq +0.6$.
The improved SEOBNRv2 model has been calibrated to
a significantly larger set of NR simulations, including eight non-spinning
simulations with $q\leq 8$ and $30$ spinning, non-precessing 
simulations~\cite{Taracchini:2013rva}. This model is capable of modeling
binaries with non-precessing component spins over the range 
$-1\leq\chi_{1,2}\leq +1$.
We refer the reader to~\citep{Taracchini:2012,Taracchini:2013rva} for a 
comprehensive summary of the technical details of these two models. 
We note that due to the high computational cost of evaluating these models,
both current LIGO searches and we use a reduced-order model of
SEOBNRv2~\cite{Purrer:2015tud} for search templates.

\subsubsection{Phenomenological}
Offline GW searches and parameter estimation efforts aimed at 
binary black holes involve filtering the detector data with modeled waveforms
in the \textit{frequency} domain. One way to minimize their computational 
cost is to use frequency-domain closed-form
GW models as search filters. Past LIGO-Virgo searches
used the TaylorF2 model (see, e.g.~\cite{Colaboration:2011np}), although
with the significant limitation that TaylorF2 describes only the inspiral
phase.
A phenomenological model (IMRPhenomC) based on it has been developed to also
capture the plunge and 
merger phase waveforms~\cite{Santamaria:2010yb}. This model uses TaylorF2
phasing and amplitude prescriptions during the early inspiral, and stitches
on an analytic ansatz for GW phasing and amplitude during 
the late-inspiral, plunge and merger phases. These ans\"{a}tze are written as
polynomials in $f^{1/3}$, where $f$ is the instantaneous gravitational-wave
frequency, and the associated coefficients are treated as free parameters.
In the ringdown regime, IMRPhenomC models binary phasing as a linear function
in $f$, capturing the effect of the leading QNM with a Lorentzian.
The model is calibrated to reproduce accurate NR waveforms
for non-precessing binaries with mass-ratios $q\leq 4$ and BH spins
between $[-0.75, +0.83]$, produced by different groups~\cite{Bruegmann2006,
Husa-Hannam-etal:2007,Pollney-Reisswig:2007,Pollney:2009yz,Scheel2006}.
The free parameters are interpolated over the binary mass and spin 
parameter space as polynomials in the symmetric mass-ratio $\eta$ and
mass-weighted spin $\chi_\mathrm{mw}$,
\begin{equation}\label{eq:chimw}
\chi_\mathrm{mw} \defeq \frac{m_1}{m_1+m_2}\chi_{1} + \frac{m_2}{m_1+m_2}\chi_{2}, 
\end{equation}
to obtain IMRPhenomC inspiral-merger-ringdown waveforms at arbitrary binary 
masses and spins.
We refer the reader to Ref.~\cite{Santamaria:2010yb} for a complete 
description of this model.

The very recent IMRPhenomD model~\cite{Khan:2015jqa} improves upon IMRPhenomC in
several crucial aspects: (i) use of both 
component spins to model the inspiral phasing, (ii) use of the 
spin parameter $\chi_\mathrm{eff}$~\cite{Ajith:2011ec},
\begin{equation}\label{eq:chieff}
 \chi_\mathrm{eff}\defeq \chi_\mathrm{mw} - \frac{38\eta}{113}(\chi_{1}+\chi_{2})
\end{equation}
(with symmetric mass-ratio $\eta=m_1 m_2/(m_1+m_2)^2$), to capture the late-inspiral/plunge
phase, (iii) use of (uncalibrated) EOB+NR hybrid waveforms to constrain 
free parameters, and (iv) use of several high mass-ratio NR simulations to extend 
the range of validity of the model. The simulations used to calibrate 
IMRPhenomD sample component spins more densely than the set used for 
IMRPhenomC, and cover mass-ratios up to $q=18$. We refer the reader to
Ref.~\cite{Khan:2015jqa,Husa:2015iqa} for further details of IMRPhenomD.


\section{Faithfulness Analysis}\label{s1:faithfulness}

\begin{figure*}
\centering
\includegraphics[width=2\columnwidth]{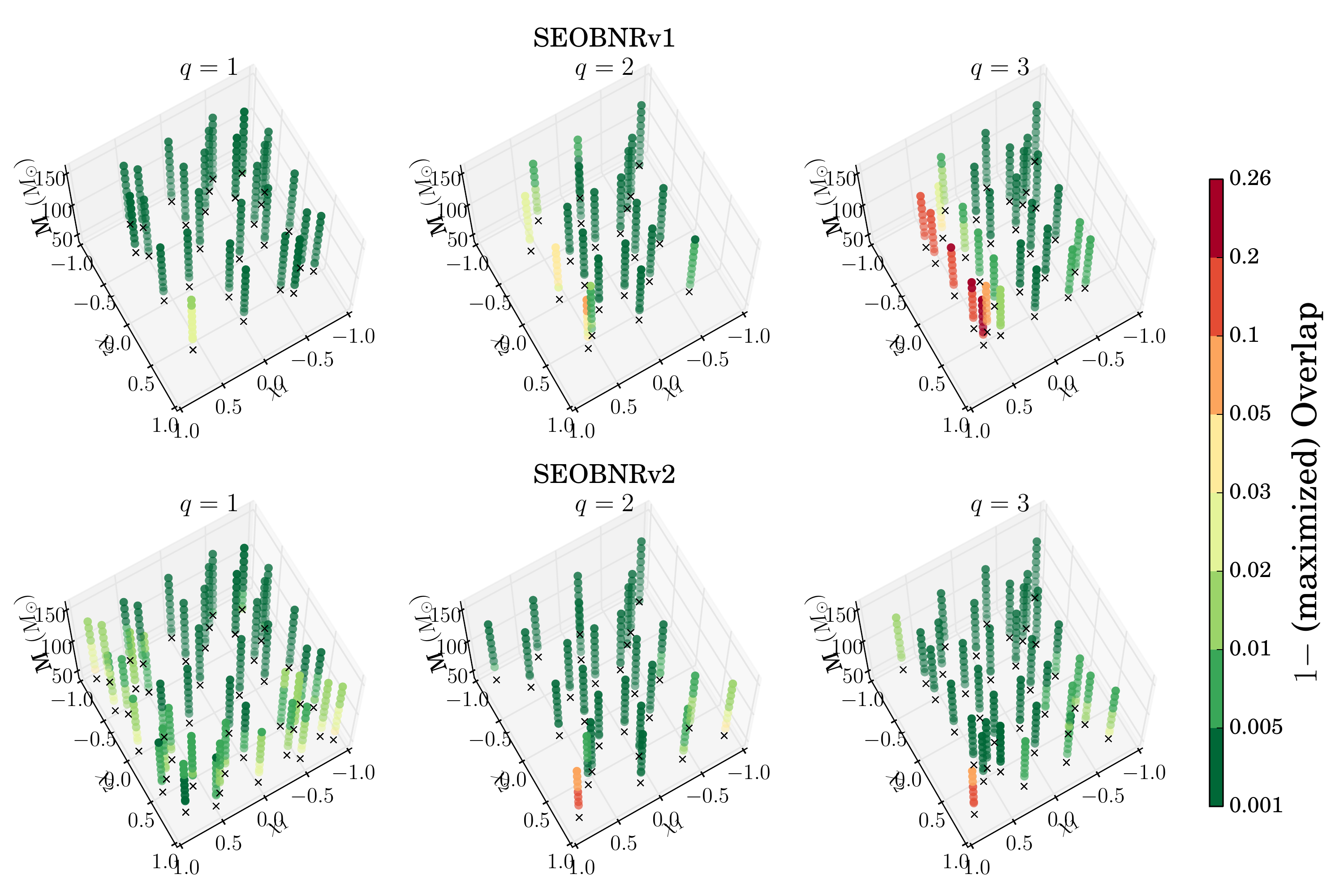}
\caption{Unfaithfulness between SEOBNR and NR waveforms as a 
function of mass-ratio $q=m_1/m_2$, component spins $\chi_{1}, \chi_{2}$, and
total mass $M$. SEOBNRv1 (top panel) reproduces NR well when the spin on the 
bigger BH does not exceed $+0.5$, with inaccuracies increasing with mass-ratio.
SEOBNRv2 (bottom panel) significantly improves over SEOBNRv1 with overlaps 
against NR higher than $98\%$ over most of the parameter space considered. 
However, when spins on both component BHs are large and positive-aligned, 
SEOBNRv2 fails to produce accurate waveforms ($\mathcal{O}\simeq 0.80$). We 
note that both models are accurate within their respective calibration range, 
but become inaccurate outside this range. Therefore it is crucial to test 
waveform models before using them in aLIGO analyses.
}
\label{fig:SEOB_unfaith_TotalMass_Spin1z_Spin2z}
\end{figure*}
\begin{figure}[h]
\centering
\includegraphics[width=1.05\columnwidth, trim={{0.05\columnwidth} 0 0 0}]{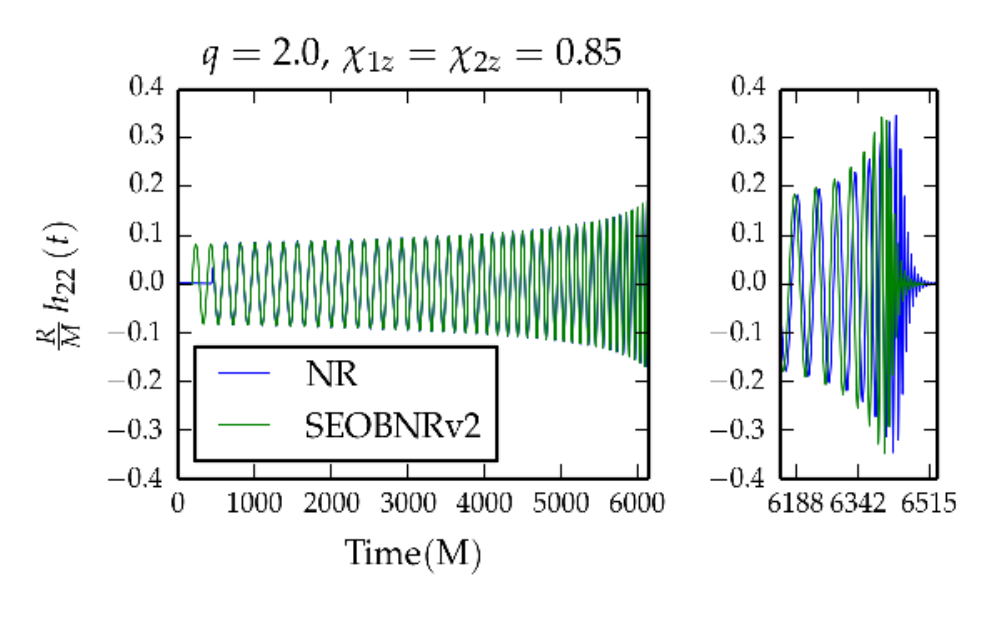}\\
\includegraphics[width=1.05\columnwidth, trim={{0.05\columnwidth} 0 0 0}]{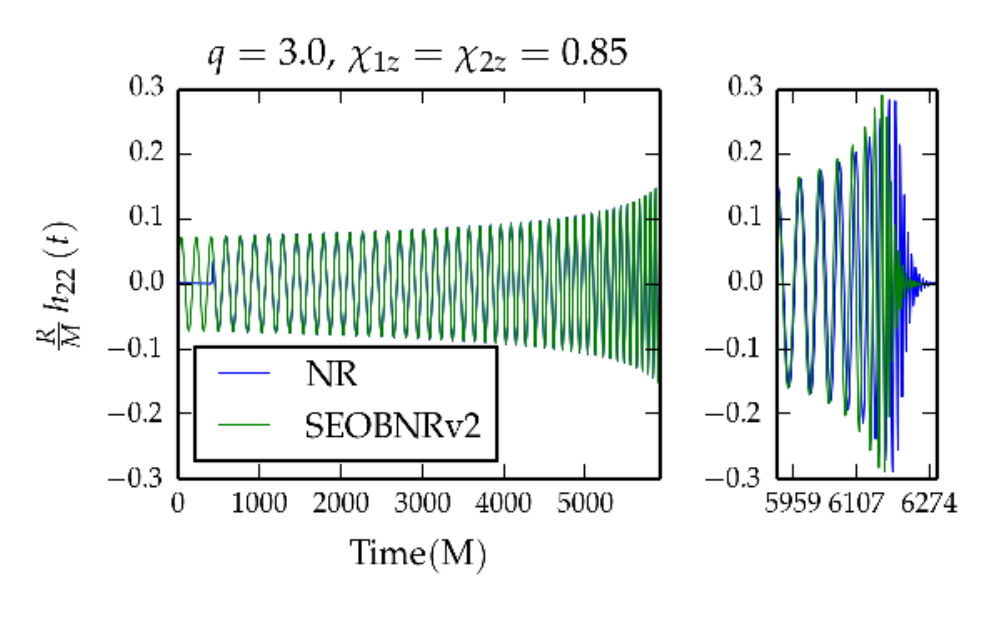}
\caption{SEOBNRv2 and NR waveforms for the problematic cases identified in the
high spin corner of Fig.~\ref{fig:SEOB_unfaith_TotalMass_Spin1z_Spin2z}.
Top: $q=2,\,\chi_1=\chi_2=+0.85$. Bottom: $q=3,\,\chi_1=\chi_2=+0.85$. 
Waveforms are aligned during their first few inspiral cycles.
}
\label{fig:SEOBv2_unfaith_q23s85s85cases}
\end{figure}
\begin{figure*}
\centering
\includegraphics[width=2\columnwidth]{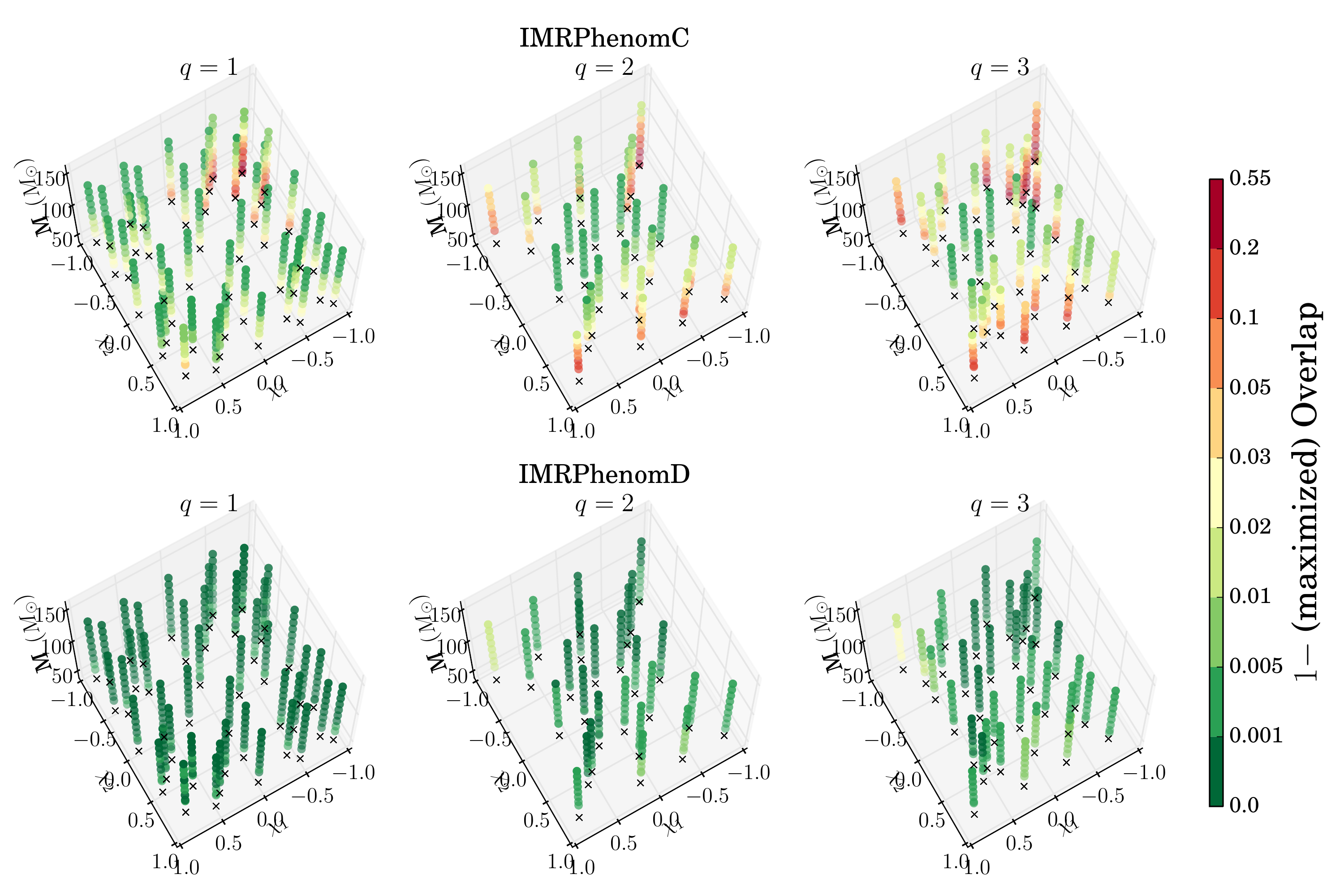}
\caption{This figure is similar to
Fig.~\ref{fig:SEOB_unfaith_TotalMass_Spin1z_Spin2z} with the difference
that the models considered here are IMRPhenomC and IMRPhenomD (top and
bottom panels, respectively). We note that both of the phenomenological
models have been calibrated over most of the mass-ratio and spin range probed here.
While IMRPhenomC shows significant deviation from NR as soon as we increase
the mass-ratio above $q=1$, and/or spin magnitudes above $\approx 0.5$,
we find that IMRPhenomD reproduces NR remarkably well with
overlaps above $99\%$ everywhere (above $99.5\%$ over most of the space).
}
\label{fig:PhenomCD_unfaith_TotalMass_Spin1z_Spin2z}
\end{figure*}
\begin{figure*}[h]
\centering
\includegraphics[width=2\columnwidth]{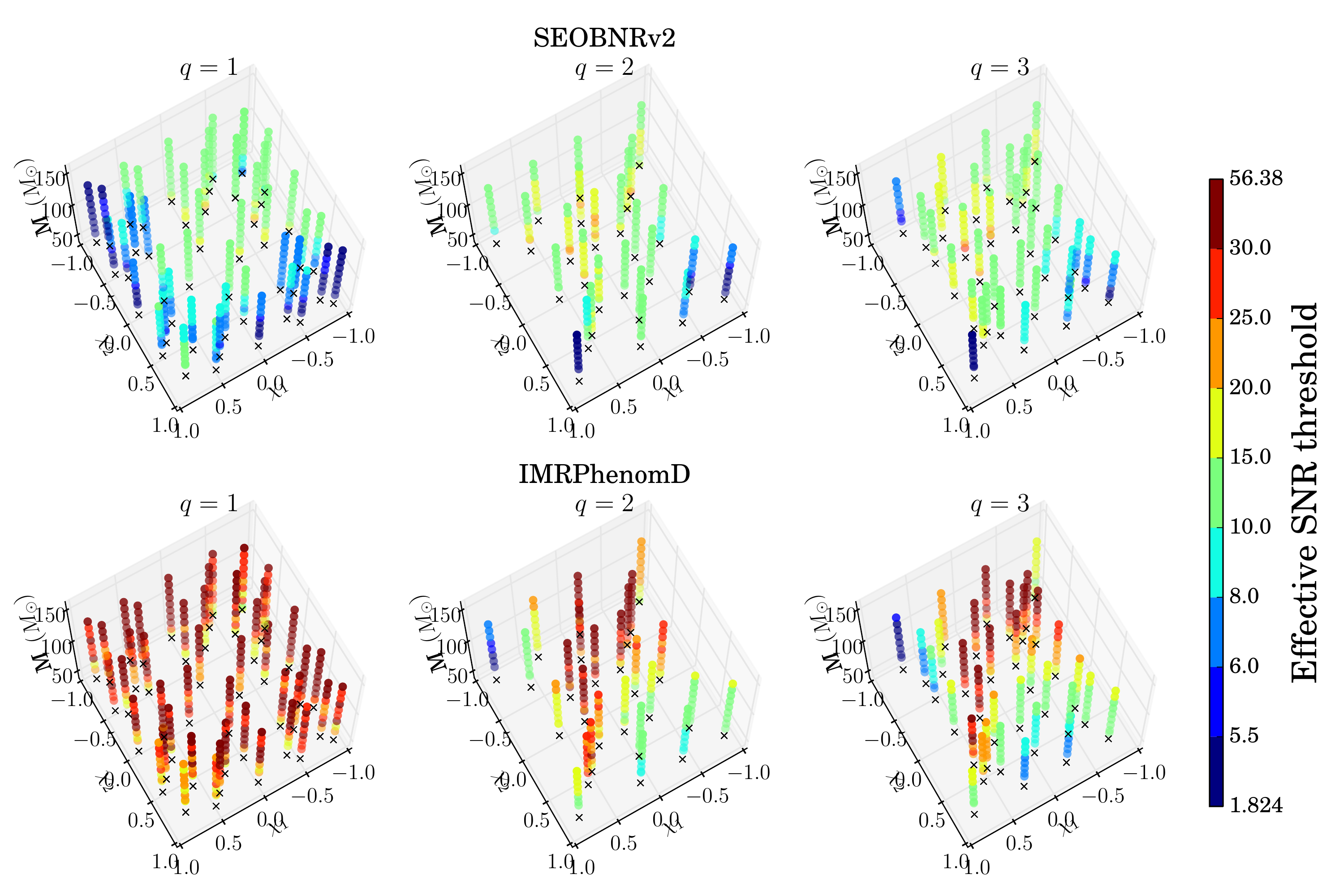}
\caption{We show the effective SNR level at which the SEOBNRv2 and IMRPhenomD
models become distinguishable from NR waveforms with the Advanced LIGO
instruments. Here we use the indistinguishability criterion proposed in 
Ref.~\cite{Lindblom2008}.
}
\label{fig:SEOB_snrEff_TotalMass_Spin1z_Spin2z}
\end{figure*}

We now proceed to a comparison of the NR waveforms introduced in 
Sec.~\ref{s2:numrel_simulations} with the analytical waveform models introduced
in Sec.~\ref{s2:waveform_models}, beginning with an analysis of their
faithfulness (c.f. Eq.~\ref{eq:overlap}).
We rescale the NR waveforms to
a range of total masses, and compute overlaps with model waveforms with 
identical BH parameters. These overlaps are maximized over 
the extrinsic parameters however, i.e. over the time and phase at coalescence.
They measure the accuracy of the models
at specific points in the parameter space $(m_1,m_2,\chi_{1},\chi_{2})$.

In Fig.~\ref{fig:SEOB_unfaith_TotalMass_Spin1z_Spin2z} we show the 
unfaithfulness (i.e. $1-\mathcal{O}$) of the two EOB models, SEOBNRv1 and
SEOBNRv2. In each row, the three
panels correspond to mass ratios $q=\{1,2,3\}$. In each panel, the three axes 
correspond to component spins and total mass with the color showing the 
unfaithfulness. 
Note that the total-masses probed are restricted to $M\gtrsim 50M_\odot$
(c.f. Fig.~\ref{fig:Min_TotalMass_Spin1z_Spin2z}).

For SEOBNRv1, we find that its unfaithfulness increases with binary mass-ratio as
well as with the more massive component's spin, with little dependence on
the binary's total mass. From the top left panel in 
Fig.~\ref{fig:SEOB_unfaith_TotalMass_Spin1z_Spin2z} we note that for the 
smallest mass-ratio, $q=1$, SEOBNRv1 reproduces the NR waveforms well with 
unfaithfulness below $0.5\%$ over most of the spin parameter space, except
when the spins on both holes are close to the maximum value that the model
supports (i.e. $+0.6$), where its unfaithfulness rises above $2\%$. As we increase 
the mass-ratio to $q=2$ (top middle panel of the same figure) SEOBNRv1's 
faithfulness further drops below $95\%$ in the high-aligned-spin region. 
Furthermore, we also find that the unfaithfulness of the model reaches $1-2\%$ 
when the smaller hole carries large anti-aligned spin. Further increasing the 
mass-ratio to $q=3$ increases the differences of the model with NR further,
with overlaps
falling below $90\%$ when the larger black hole's spin $\rightarrow +0.6$.
Overall, we find that the model performs better when the more massive hole
has anti-aligned spins rather than aligned.

Turning to the SEOBNRv2 model, we find that it significantly improves over
SEOBNRv1: for equal-mass binaries, we find from the bottom left panel of
Fig.~\ref{fig:SEOB_unfaith_TotalMass_Spin1z_Spin2z} that the unfaithfulness of
SEOBNRv2 is generally better than $1\%$ except for mixed aligned/anti-aligned
spin directions of large spin-magnitudes, where its unfaithfulness reaches $3\%$.
For higher mass-ratios $q=\{2,3\}$, the slight increase of unfaithfulness 
towards the aligned/anti-aligned spin corner persists. For instance, 
$1-\mathcal{O}\simeq 0.97$ for $q=2, \chi_{1}=-0.85, \chi_{2}=+0.85$.
However, the most significant deviation between SEOBNRv2 and NR occurs for both
spins aligned with large magnitudes. For $\chi_{1}=\chi_{2}=+0.85$, the
unfaithfulness rises above $10\%$ for mass-ratios $q=\{2,3\}$.
We explore these differences between SEOBNRv2 and NR further.
In Fig.~\ref{fig:SEOBv2_unfaith_q23s85s85cases}, we compare the model and NR 
waveforms for $q=\{2,3\},\,\chi_1=\chi_2=+0.85$. In both panels, the waveform
pairs are aligned near the start of the NR waveform. We find that 
the SEOBNRv2 phase evolution agrees with NR during most of the inspiral phase, 
but its frequency rises faster during the plunge phase than found with NR,
resulting in an artificially accelerated merger. This evidence hints that the
calibration of 
the merger portion and ringdown attachment of SEOBNRv2 will need further tuning.

We now turn our attention to the phenomenological models IMRPhenomC/D. 
The unfaithfulness of IMRPhenomC and IMRPhenomD with respect to NR, shown in
Fig.~\ref{fig:PhenomCD_unfaith_TotalMass_Spin1z_Spin2z}, displays patterns
distinct from the SEOBNR models. 
We find that IMRPhenomC shows poorer agreement with NR than either of the
SEOBNR models, with unfaithfulness \textit{increasing} rapidly with 
\textit{mass-ratio}, \textit{spin magnitudes}, and with 
\textit{decreasing binary masses}.
The top panels of Fig.~\ref{fig:PhenomCD_unfaith_TotalMass_Spin1z_Spin2z}
show that this disagreement rises to $10-15\%$ unfaithfulness, especially
as the spin magnitude of the smaller BH grows.
we notice disagreement between PhenomC and NR for large anti-aligned spins, 
which increases to $10-15\%$ unfaithfulness over most of the spin parameter 
space as we go from $q=1$ to $q=\{2,3\}$. This disagreement increases, also,
as more of the NR waveform is integrated over, i.e. at lower masses.
In stark contrast, the newest model considered, IMRPhenomD, shows better 
agreement with NR than either of the SEOBNR models, with faithfulness above
$99\%$ over most of the analyzed parameter space, as seen in the bottom 
panels of Fig.~\ref{fig:PhenomCD_unfaith_TotalMass_Spin1z_Spin2z}. The only 
region where we see somewhat smaller overlaps is for $q\ne 1$ mixed-aligned 
spins with large positive spin on the larger hole.

We conclude that both SEOBNRv2 and IMRPhenomD models describe well binaries 
with \textit{low to moderate spins}, and even \textit{high anti-aligned spins},
with the latter also representing well \textit{high aligned spins} binaries.
The accuracy of both degrades somewhat with increasing mass-ratio in the  
high aligned/aligned spin and high aligned/anti-aligned spin corners of the 
parameter space respectively. We also find that both of
these models outperform their earlier counterparts significantly.

Further, we ask the question: how loud does a GW signal have to be for modeling
errors to degrade scientific conclusions derived from it.
To answer that, we use the 
sufficient criterion $(\delta h|\delta h)< 1$, where 
$\delta h=h^\mathrm{true}-h^\mathrm{modeled}$, to calculate the SNR threshold
$\rho_\mathrm{eff}$
below which the true and modeled waveforms will not be distinguishable by 
aLIGO~\cite{Lindblom2008}, i.e.
\begin{equation}\label{eq:LindbloomOwenBrownCriterion}
 \rho_\mathrm{eff} = \dfrac{1}{\sqrt{2(1-\mathcal{O}(h^\mathrm{NR}, h^\mathrm{modeled}))}}.
\end{equation}
$\rho_\mathrm{eff}$ is the threshold value of the GW SNR, such that for 
$\rho\leq\rho_\mathrm{eff}$ the statistical errors in mass and spin estimation
will
dominate over any systematic biases due to model inaccuracies, and therefore
our scientific conclusions will not be degraded by model choice. 
The condition $\rho\leq \rho_\mathrm{eff}$ is necessary, but not sufficient, i.e.  it is not necessarily true that
for all $\rho\ge\rho_\mathrm{eff}$ modeling inaccuracies will actually 
dominate~\cite{Lindblom2008}. With this caveat, we show in 
Fig.~\ref{fig:SEOB_snrEff_TotalMass_Spin1z_Spin2z} the SNR threshold 
$\rho_\mathrm{eff}$
for the SEOBNRv2 and IMRPhenomD models, as a 
function of binary mass-ratio, total mass, and component spins.
From the top row of the figure, we find that SEOBNRv2 is sufficiently 
accurate for all aLIGO measurement purposes when concerned with moderately
spinning binaries at SNRs up to $\approx 15-20$. However, (i) for 
equal-mass binaries with \textit{large mixed-aligned} spins, and (ii) for 
unequal-mass binaries with \textit{large aligned} spins, using SEOBNRv2 
waveforms may lead to loss in information at fairly low aLIGO SNRs.

Turning to IMRPhenomD (lower panels of
  Fig.~\ref{fig:SEOB_snrEff_TotalMass_Spin1z_Spin2z}), we observe that
  this model is particularly accurate for equal-mass and/or equal-spin
  binaries and will be indistinguishable from NR for SNRs
  up to $\approx 30$, possibly even higher\footnote{ The agreement between IMRPhenomD and the NR waveforms is so good, that NR error estimates are of comparable order.}. The SNR-threshold falls to $\approx 15$ for
  unequal-mass unequal-spin systems. Overall, we find that IMRPhenomD
  is best suited for aLIGO parameter estimation efforts aimed at
  comparable mass-ratio aligned-spin binaries of high total mass
  ($M\geq M_\mathrm{min}\gtrsim 50M_\odot$).

\section{Effectualness}\label{s1:effectualness}
%
\begin{figure*}
\centering    
\includegraphics[width=1.9\columnwidth]{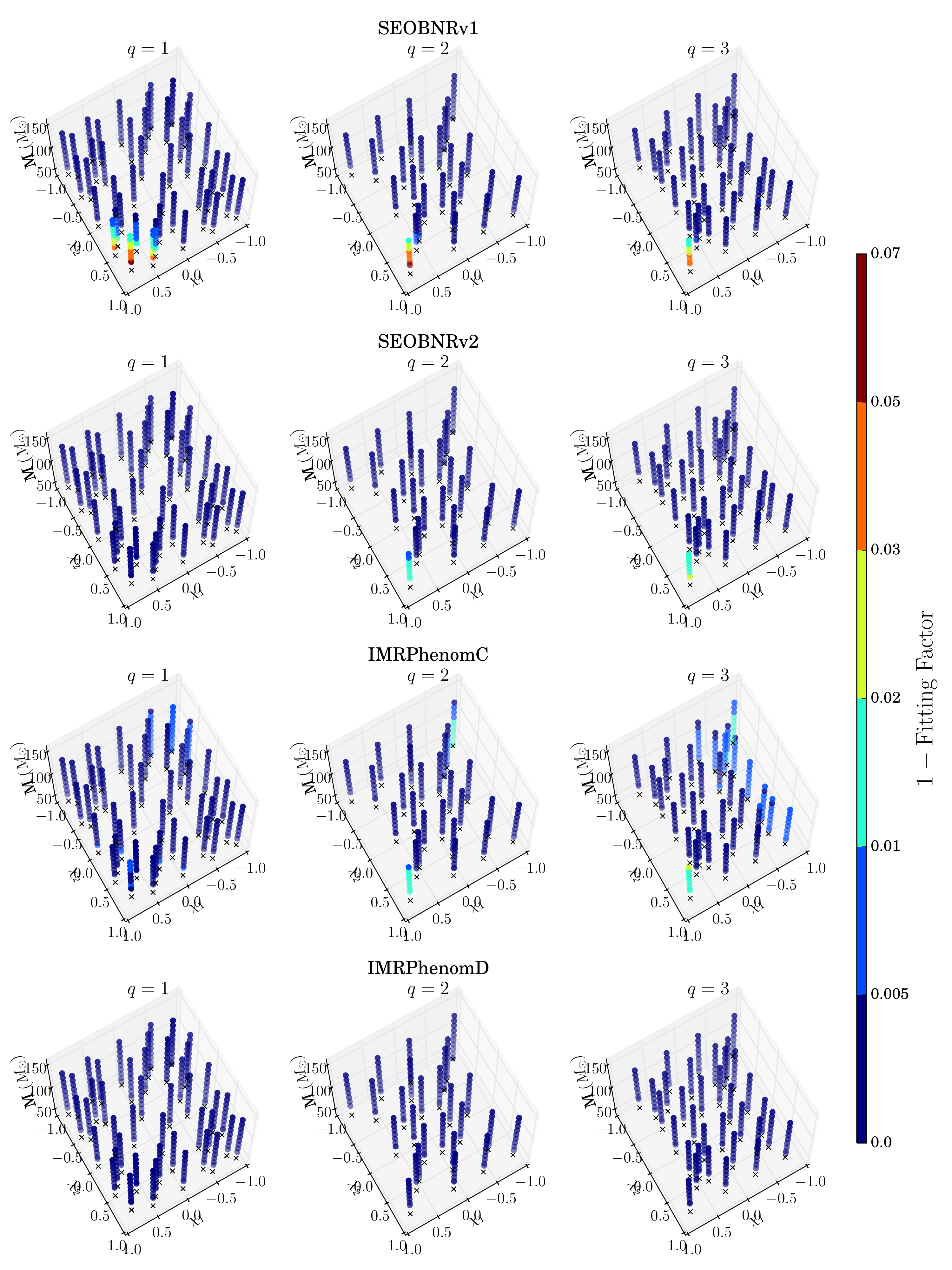}
\caption{%
Effectualness of the four waveform models considered. Plotted is the 
fractional loss in recovered SNR. Rows correspond to different models,
and within each row, the data is plotting as a function of mass-ratio $q$,
BH spins $\chi_1, \chi_2$, and total-mass $M$.
The black crosses denote the values of component spins in the $x-y$ plane.
We note that SEOBNRv1 does not model binaries with component spins higher 
than $+0.6$. We find that the SEOBNRv2 and IMRPhenomD models 
supersede their earlier counterparts, SEOBNRv1 and IMRPhenomC, respectively,
with FFs over $99.5\%$ over most of the spin and mass parameter space 
probed. However, we do find that for binaries high spins on both BHs, 
IMRPhenomD clearly out-performs all others with FFs $>99.5\%$, while
SEOBNRv2's FFs against NR deteriorate to $97\%$.
}
\label{fig:combined_mismatches_TotalMass_Spin1z_Spin2z_IMR}
\end{figure*}

Matched-filtering based GW searches use modeled waveforms as waveforms to 
filter detector data and recover signals that are otherwise buried in instrument
noise. In such a search, the recovered SNR for a given signal is optimized over a discrete grid
of binary mass and spin parameters that describe the waveforms, and is 
the highest when the filter waveform matches the signal exactly.
In any real search, some fraction of the optimal SNR
is lost due to two reasons: (i) the discreteness of the set of filter waveforms, 
and (ii) inaccuracies in the modeled waveforms.
In this section we investigate the second factor for different waveform models
from the perspective of aLIGO detection searches, focusing on non-precessing 
BBHs. We use an over-dense sampling of the waveform parameter space 
to mitigate any SNR losses due to reason (i). 
For each analytical waveform family, we compute
overlaps between waveforms at all of the sampled points and with
each of our NR waveforms. For each NR waveform, the highest overlap 
yields the fraction of optimal SNR recoverable by each waveform models.

This calculation involves a maximization over physical parameters of the 
model waveforms, and is therefore computationally more expensive than the 
faithfulness comparisons, of Sec.~\ref{s1:faithfulness}.
The results of this effectualness study are summarized in 
Fig.~\ref{fig:combined_mismatches_TotalMass_Spin1z_Spin2z_IMR}. This figure
shows the ineffectualness
$\mathcal{M}\defeq1-\Eff$ (c.f. Eq.~\ref{eq:fittingfactor}) of all IMR
models considered here. From top to bottom, 
different rows correspond to SEOBNRv1, SEOBNRv2, IMRPhenomC and 
IMRPhenomD, respectively. In each row, different panels correspond
to different mass-ratios, and each panel spans the $3-$D subspace of binary
total mass + component spins.
From the top row, we immediately notice that even though SEOBNRv1 has support 
only for binaries with $\chi_{1,2}\leq +0.6$, it recovers $\geq 99.5\%$
of the optimal SNR for most of the parameter space where \textit{either}
$\chi_1\geq +0.6$ \textit{or} $\chi_2\geq +0.6$. However, when both spins are large
and aligned, its SNR recovery deteriorate to $93-95\%$.
From the second row we notice that SEOBNRv2 performs significantly better with
$\mathcal{E}^\mathrm{SEOBNRv2}\geq 99.5\%$
over most of the parameter space for
all mass-ratios considered. The recovered SNR by SEOBNRv2 drops, however, 
when both holes have large aligned spins. For $\chi_1=\chi_2=+0.85$, only 
$97\%$
of optimal SNR are recovered, with worse performance at higher mass-ratios.
From the third row, we observe that IMRPhenomC achieves better than $98\%$ SNR
recovery over the parameter space considered. When the magnitude of the spins
on \textit{both} BHs is large and they are parallel
(i.e. either both spins aligned, or both anti-aligned), the SNR loss increases
$2\%$ with increasing mass-ratio. 
By comparing with the top row of
Fig.~\ref{fig:PhenomCD_unfaith_TotalMass_Spin1z_Spin2z} we see a clear 
demonstration of how well IMRPhenomC exploits the degeneracies of the binary 
parameter space through its use of an effective spin parameter.
These results are consistent with the understanding that it was constructed 
with the aim of being an \textit{effectual} model, and calibrated in the region of the
parameter space which we probe here~\cite{Santamaria:2010yb}.
The bottom row of Fig.\ref{fig:combined_mismatches_TotalMass_Spin1z_Spin2z_IMR},
finally, shows results for IMRPhenomD.
As expected from its faithfulness measurements stated in the previous
section, this model recovers $\geq 99.5\%$ of the optimal SNR in all of the 
parameter space which we probe here. Note that this includes all high-spin/high-spin
corners, which were problematic with the other IMR waveform models.

To summarize, we find here that IMRPhenomD is the most effectual for
BH binaries with $1\leq q\leq 3$, $-0.85\leq\chi_{1,2}\leq +0.85$ and 
total masses greater than those shown in Fig.~\ref{fig:Min_TotalMass_Spin1z_Spin2z}. 
SEOBNRv2 also shows comparable fitting factor, except for a slight drop in 
SNR recovery in the high-spin/high-spin corner of the non-precessing BBH space.

\section{Systematic Parameter Biases}\label{s1:param_bias}
%
\begin{figure*}
\centering    
\includegraphics[width=\columnwidth]{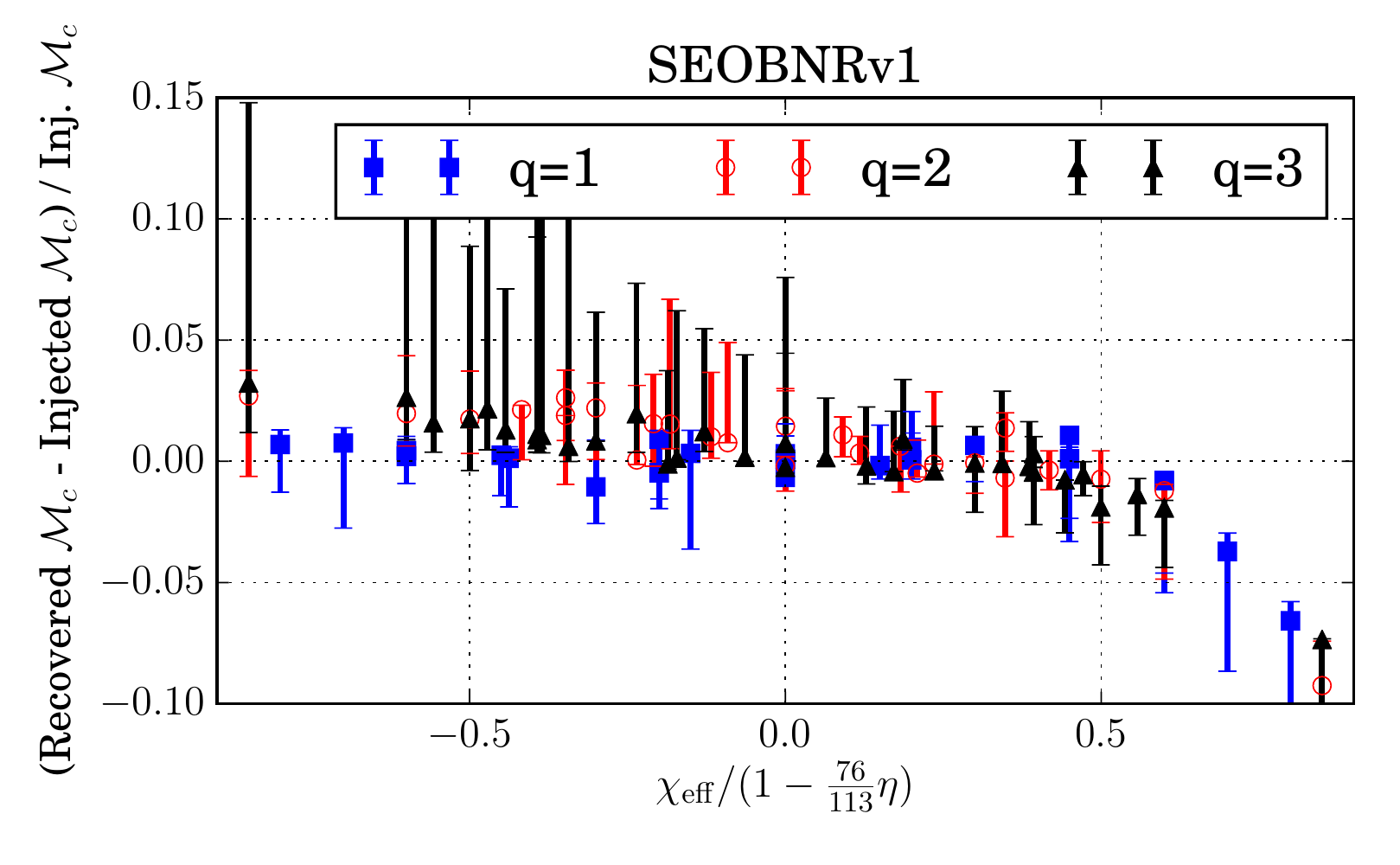}
\includegraphics[width=\columnwidth]{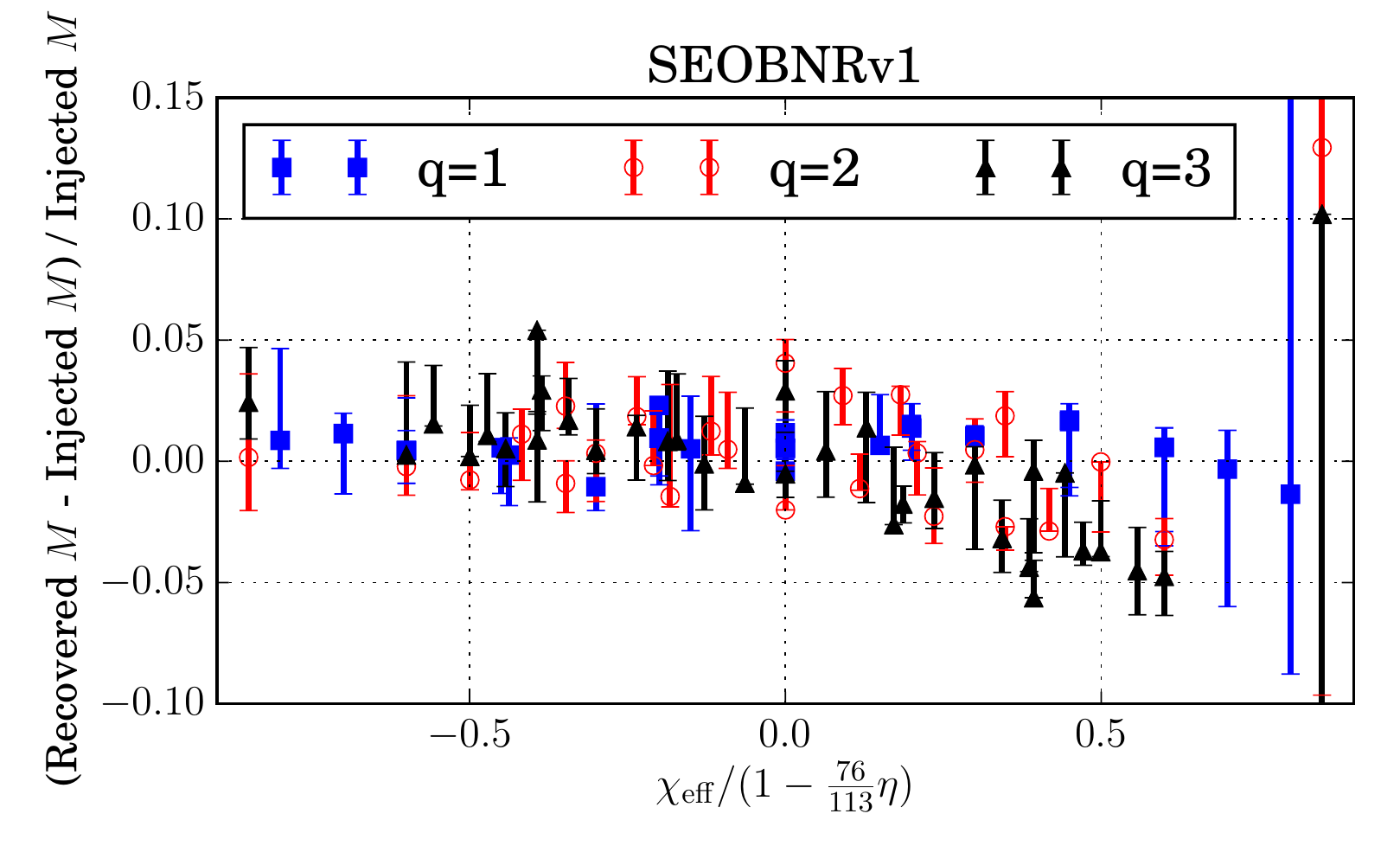}\\
\includegraphics[width=\columnwidth]{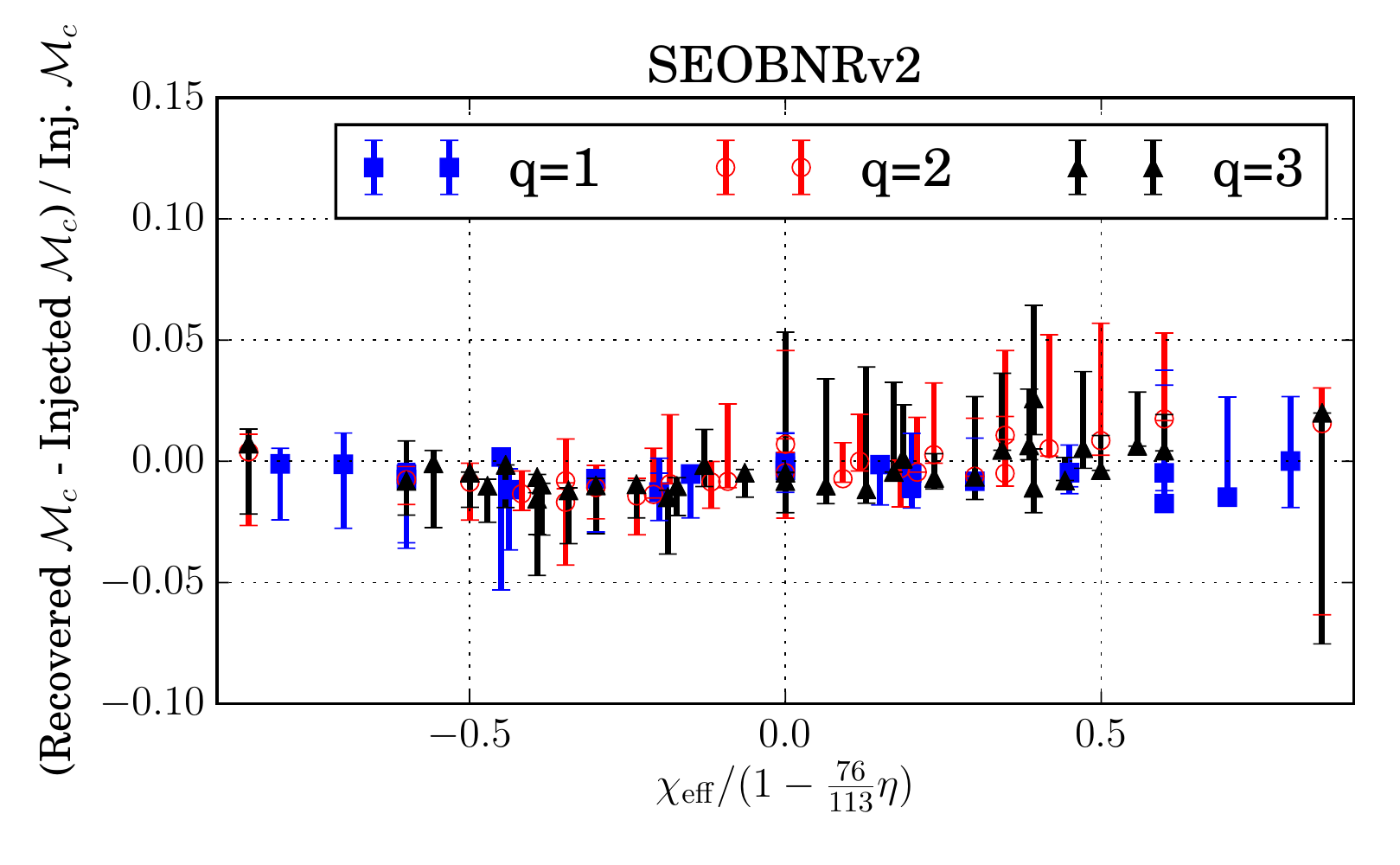}
\includegraphics[width=\columnwidth]{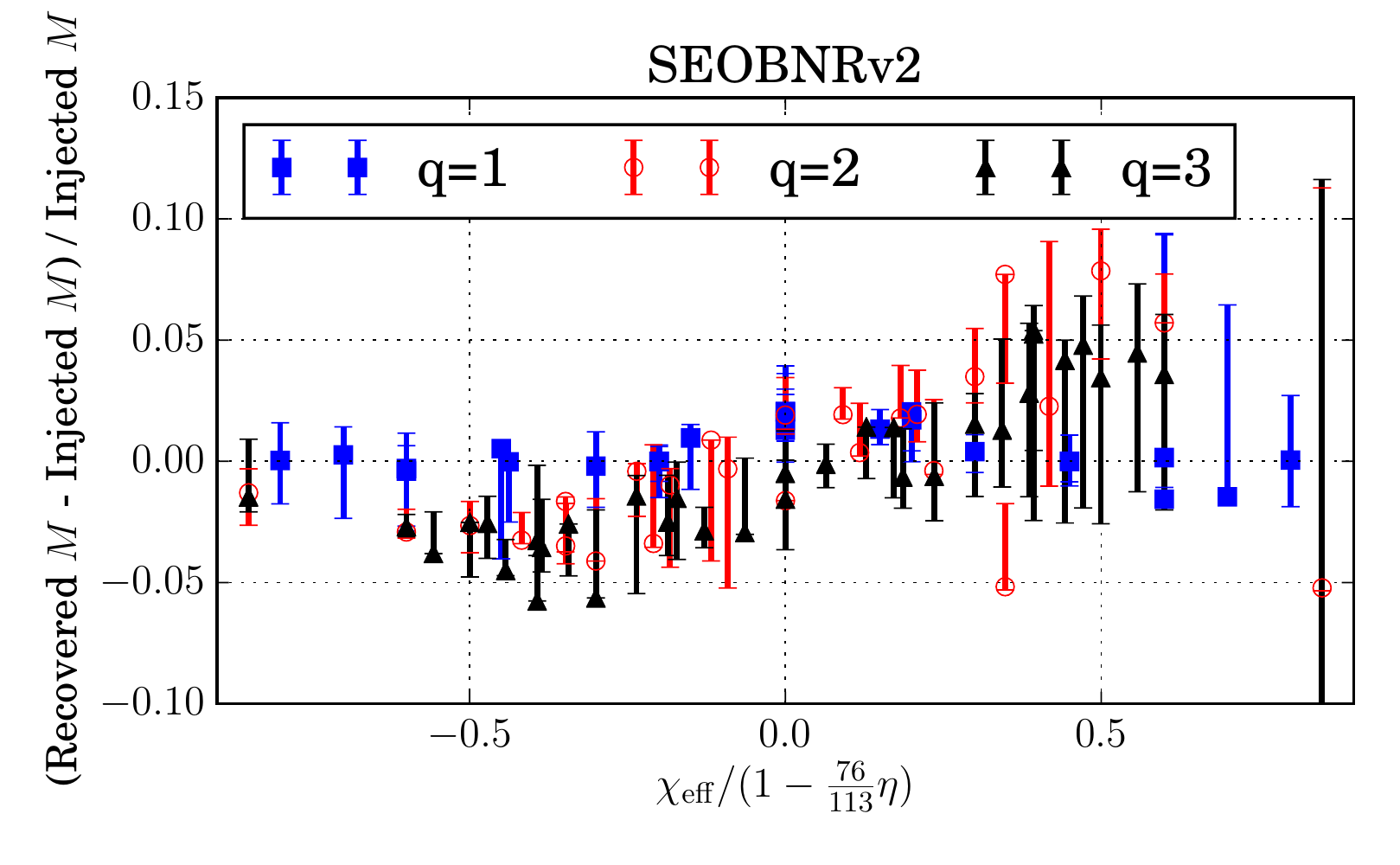}\\
\includegraphics[width=\columnwidth]{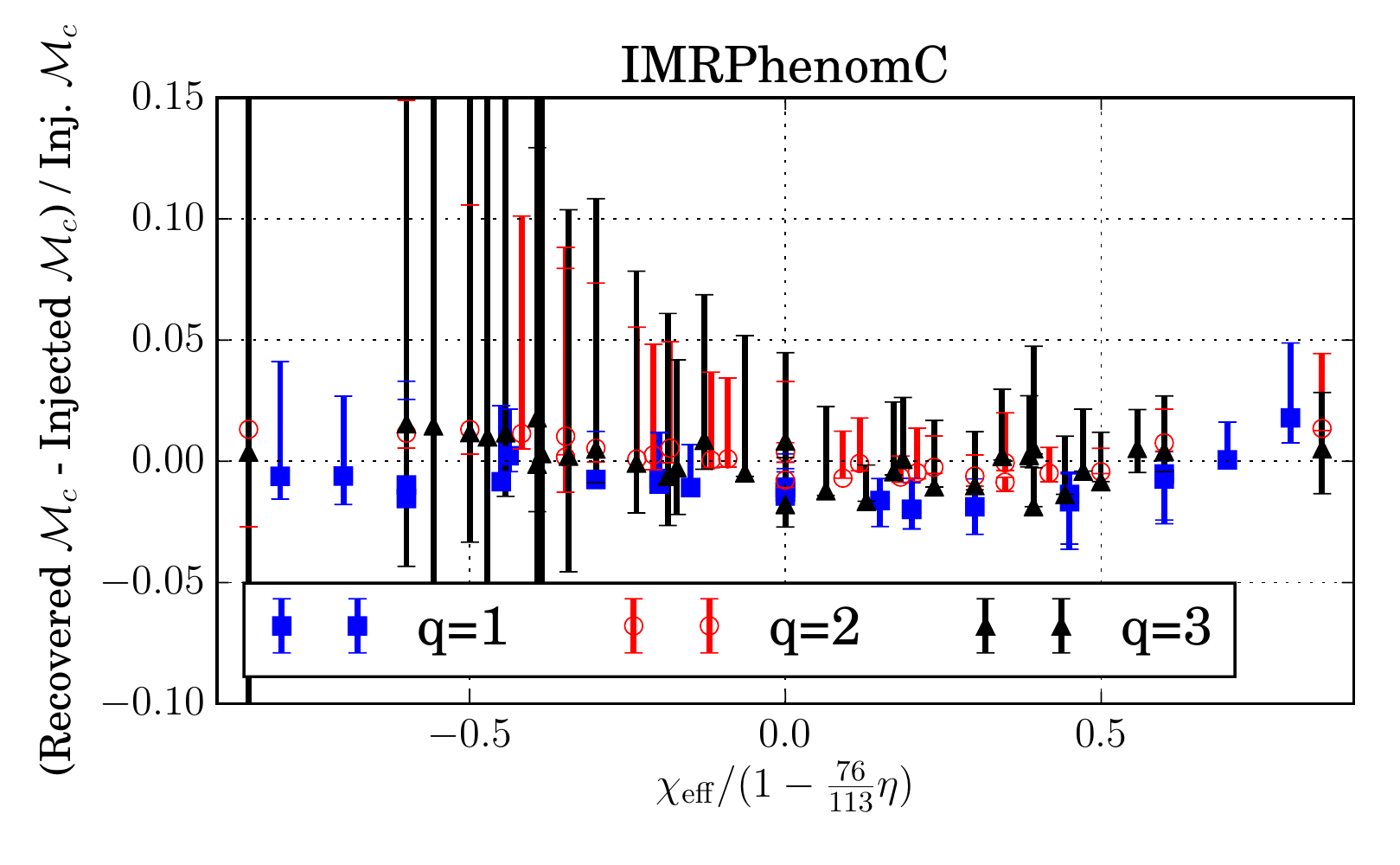}
\includegraphics[width=\columnwidth]{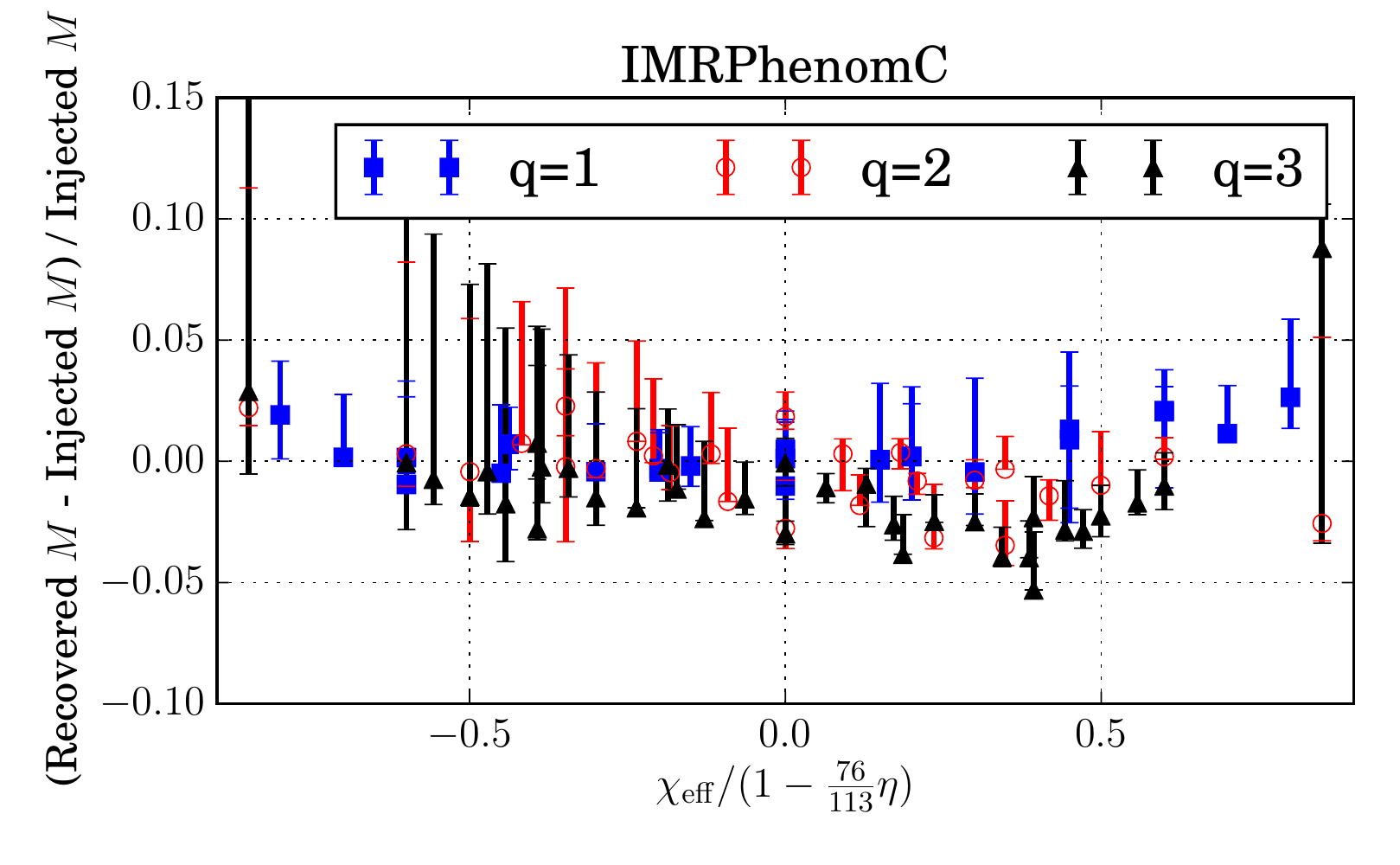}\\
\includegraphics[width=\columnwidth]{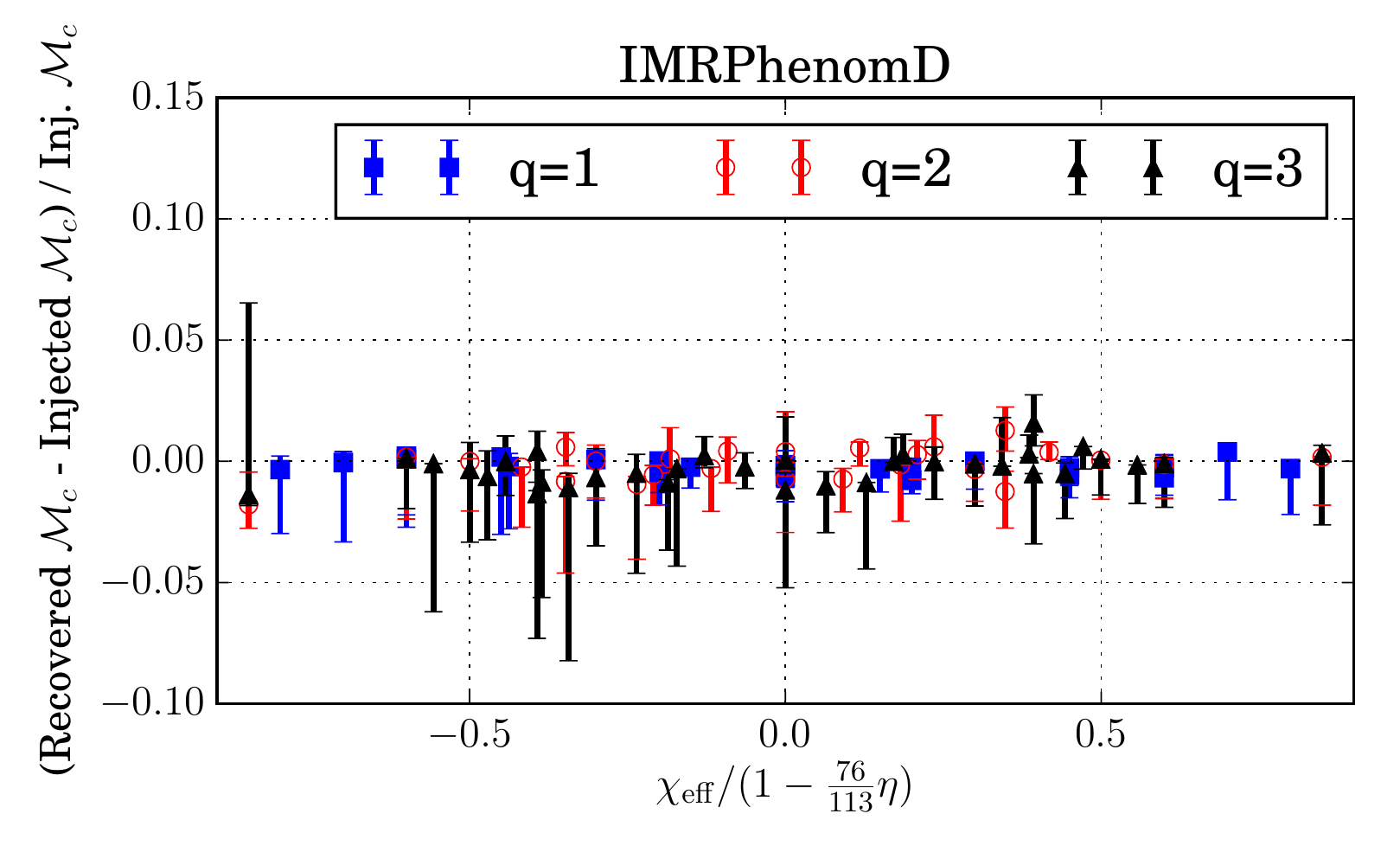}
\includegraphics[width=\columnwidth]{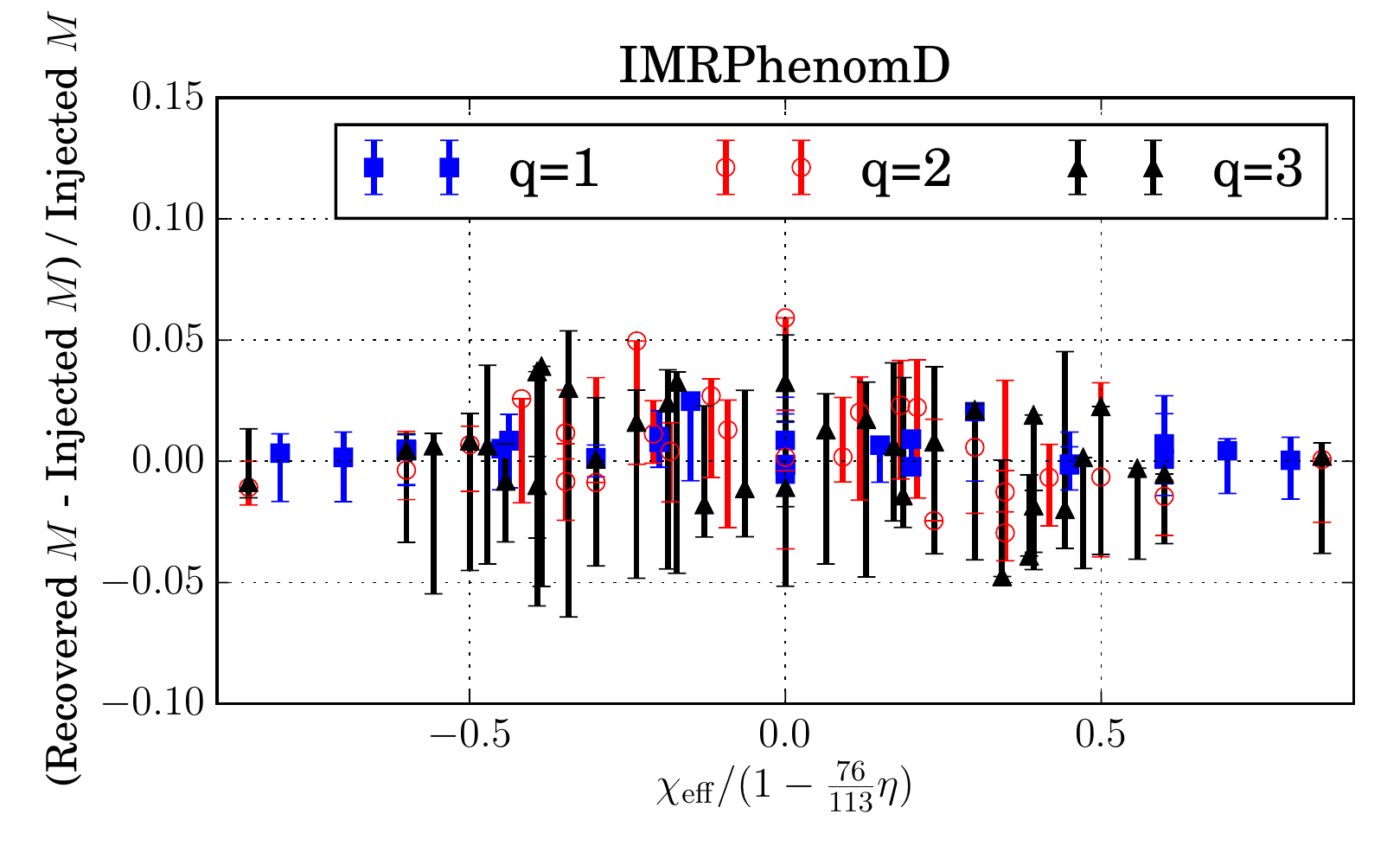}
\caption{
Systematic bias in the recovery of chirp mass $\mchirp$ (left column), and
total mass $M$ (right column) for different waveform models (rows). In each
panel, the respective bias is shown as a function of the normalized 
effective spin of the NR waveforms. The plot-markers show the bias
for a binary with total mass fixed at $M=80M_\odot$. The plot-markers
show the bias for a binary with total mass fixed at $80M_\odot$. The ``error-bars''
show the range of biases for total masses between the minimum allowed
mass and $150M_\odot$.
}
\label{fig:IMR_ChirpMassError_vs_ChiEff}
\end{figure*}
\begin{figure*}
\centering    
\includegraphics[width=\columnwidth]{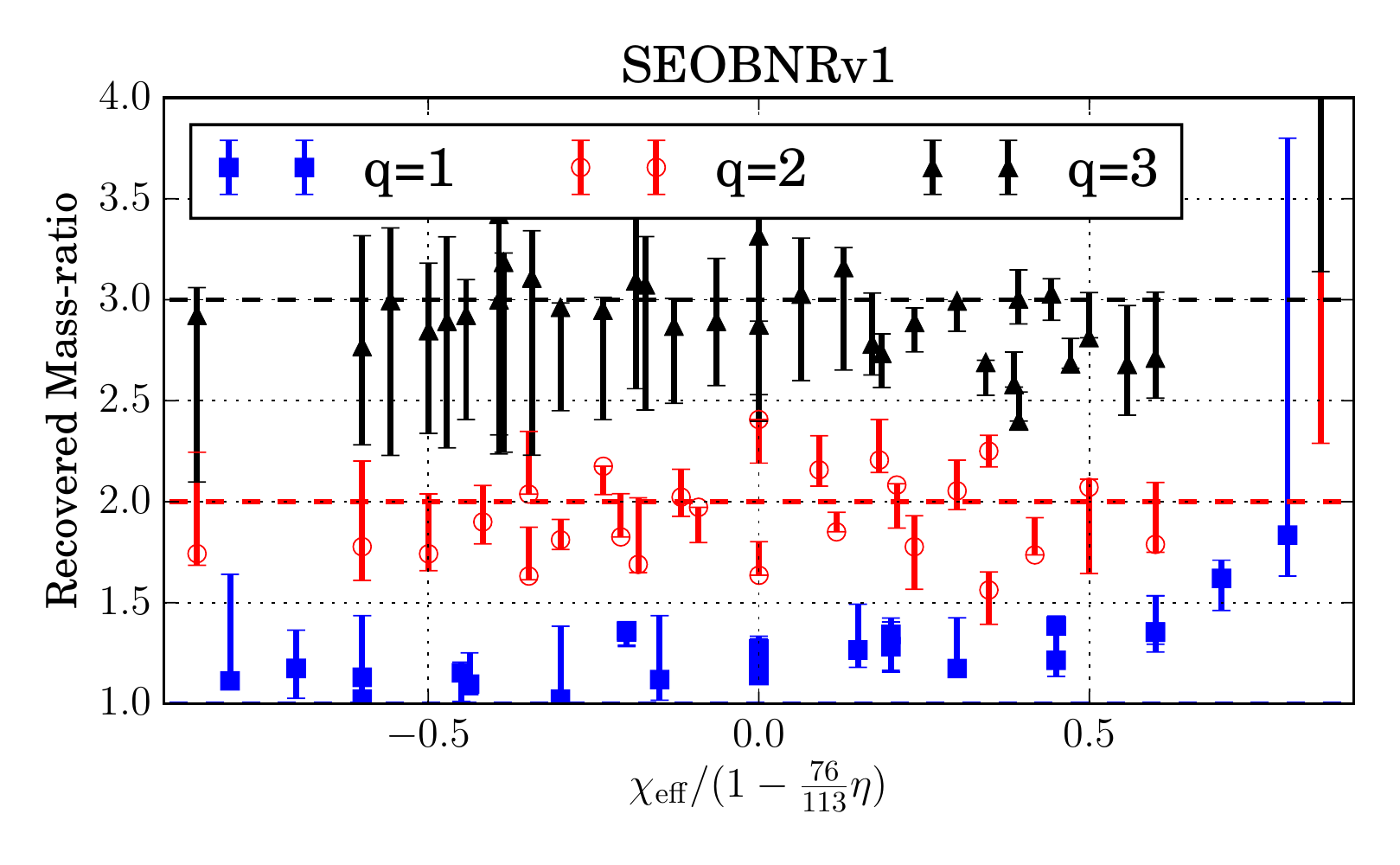}
\includegraphics[width=\columnwidth]{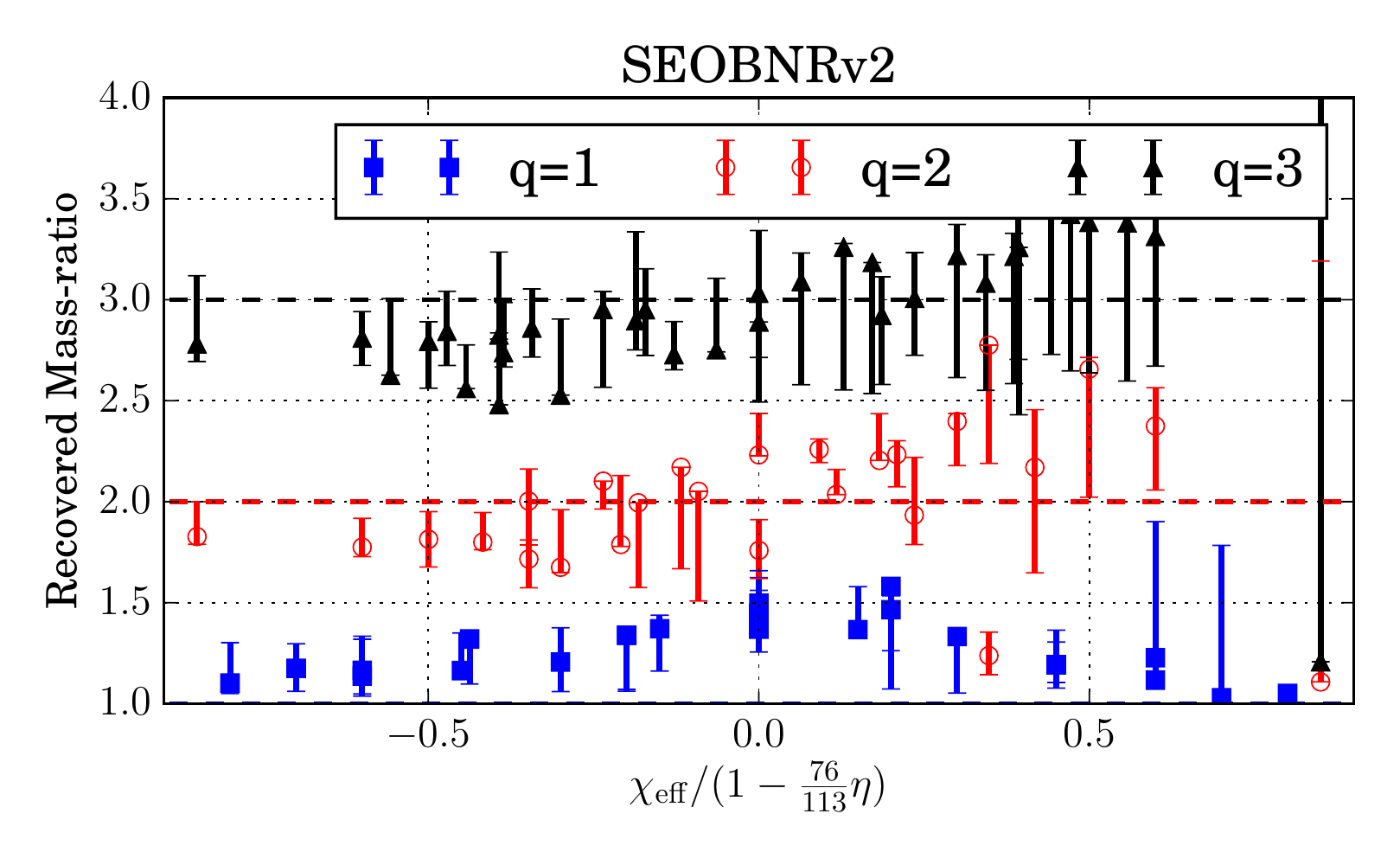}\\
\includegraphics[width=\columnwidth]{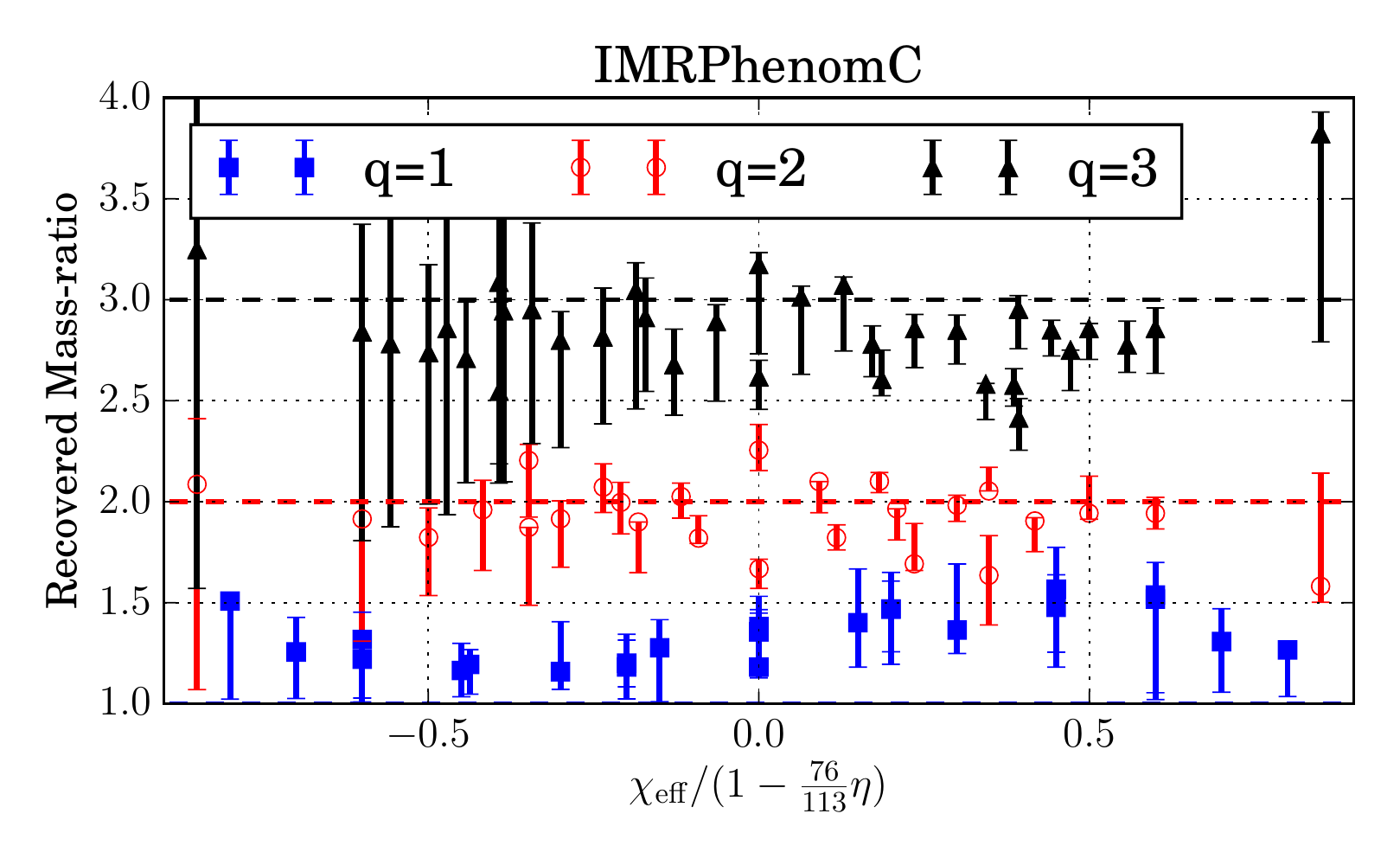}
\includegraphics[width=\columnwidth]{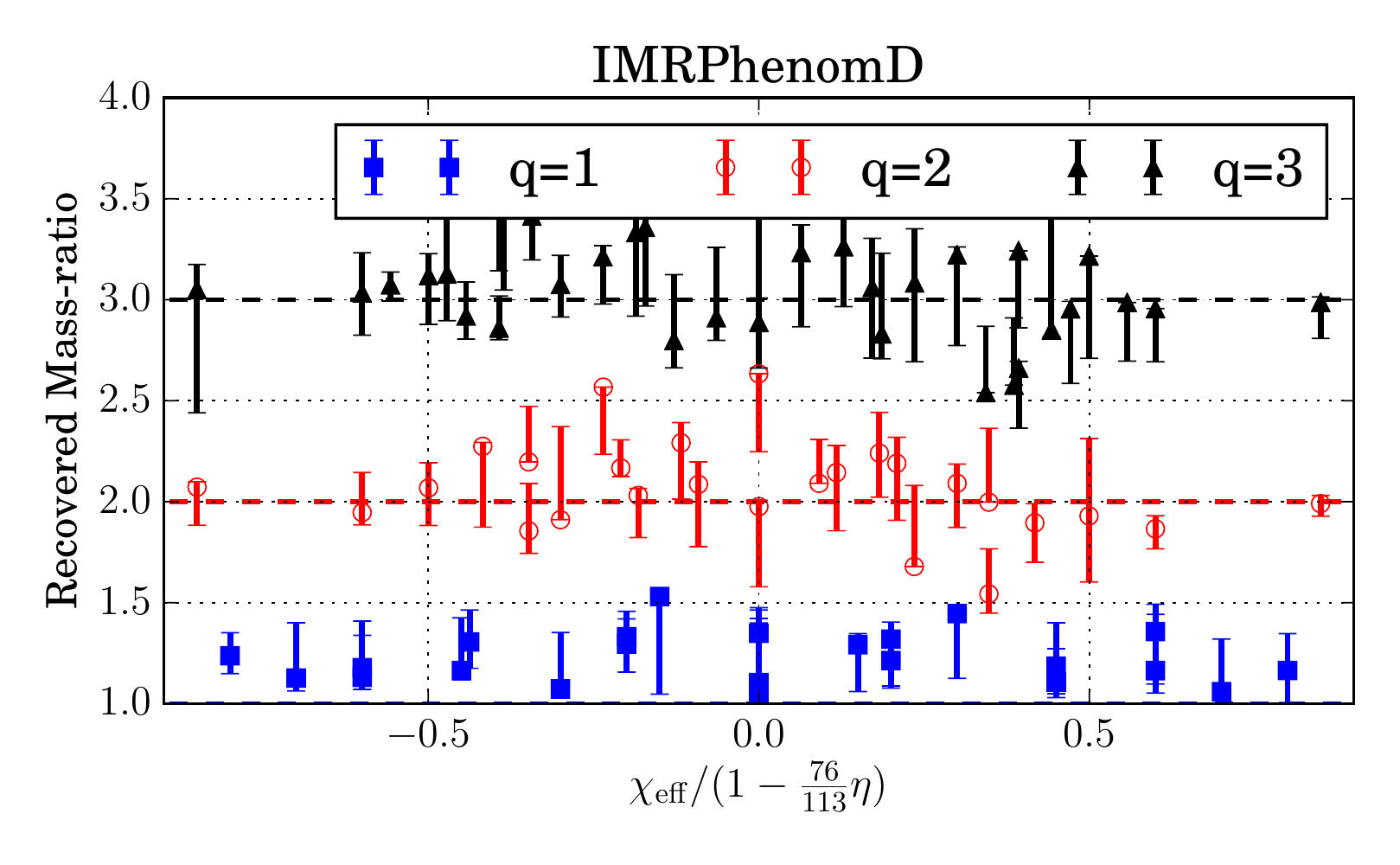}
\caption{Systematic bias in the recovery of the binary mass-ratio 
$q\defeq m_1/m_2$, as a function of the normalized effective spin of the NR
waveforms. Different mass-ratios are shown with different color, with 
horizontal dashed lines of the same color drawn to guide the eye. The plot-markers
show the recovered $q$ for a binary with total mass fixed at $80M_\odot$, while
the ``error-bars'' show the range spanned by the recovered $q$ as the 
injected binary mass is varied between its lowest allowed value and 
$150M_\odot$.
}
\label{fig:IMR_EtaRec_vs_ChiEff}
\end{figure*}
\begin{figure*}
\centering    
\includegraphics[width=\columnwidth]{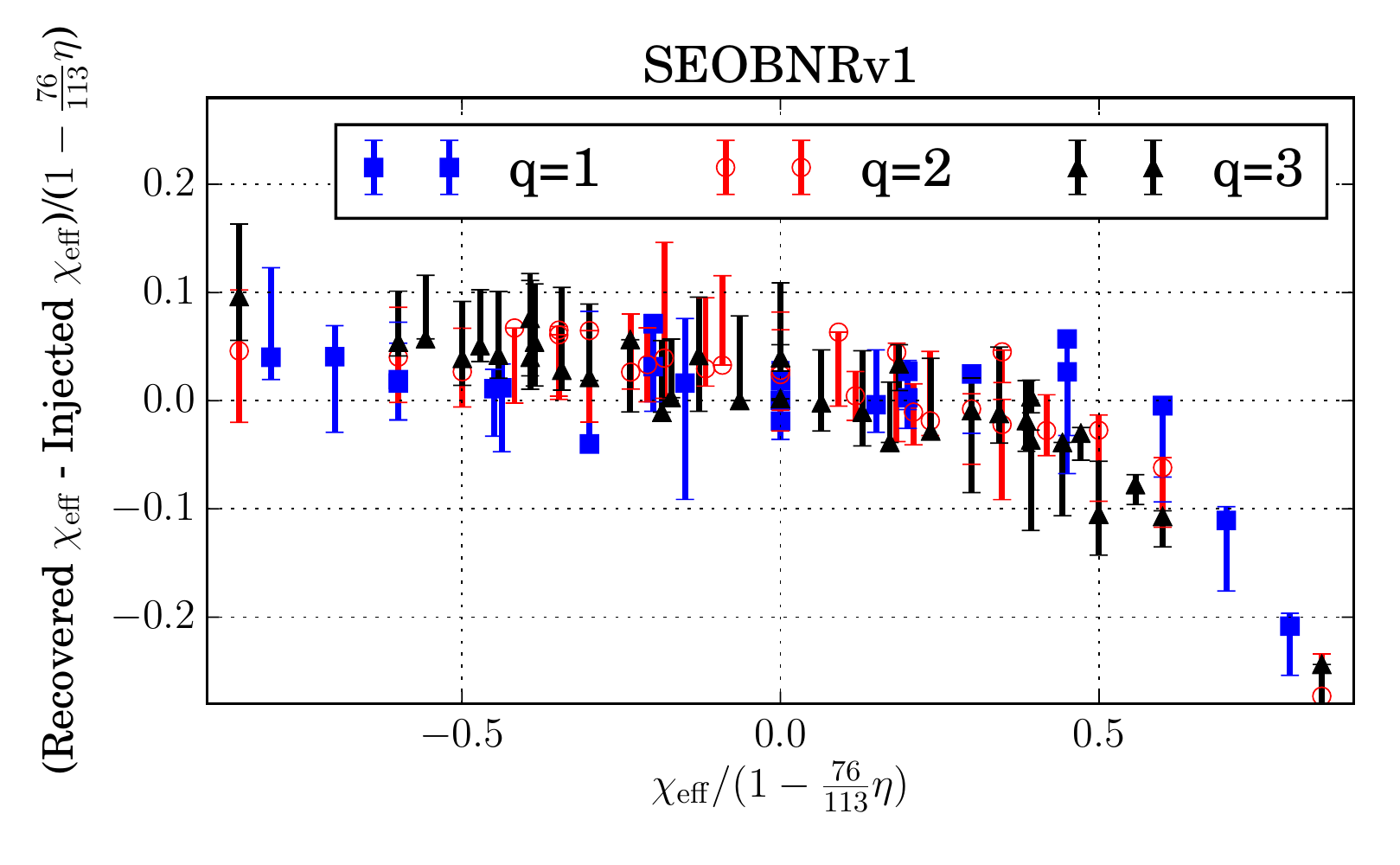}
\includegraphics[width=\columnwidth]{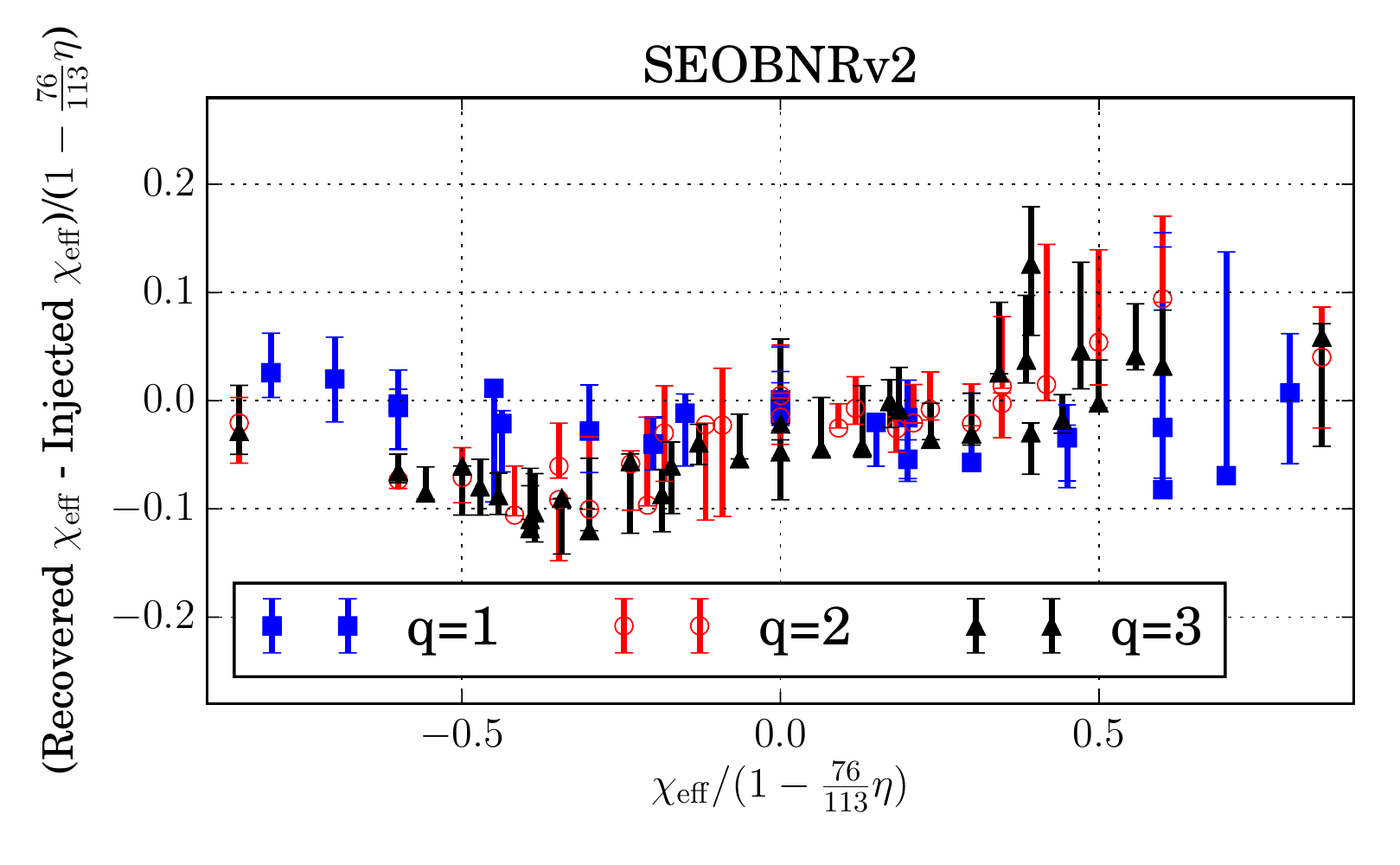}\\
\includegraphics[width=\columnwidth]{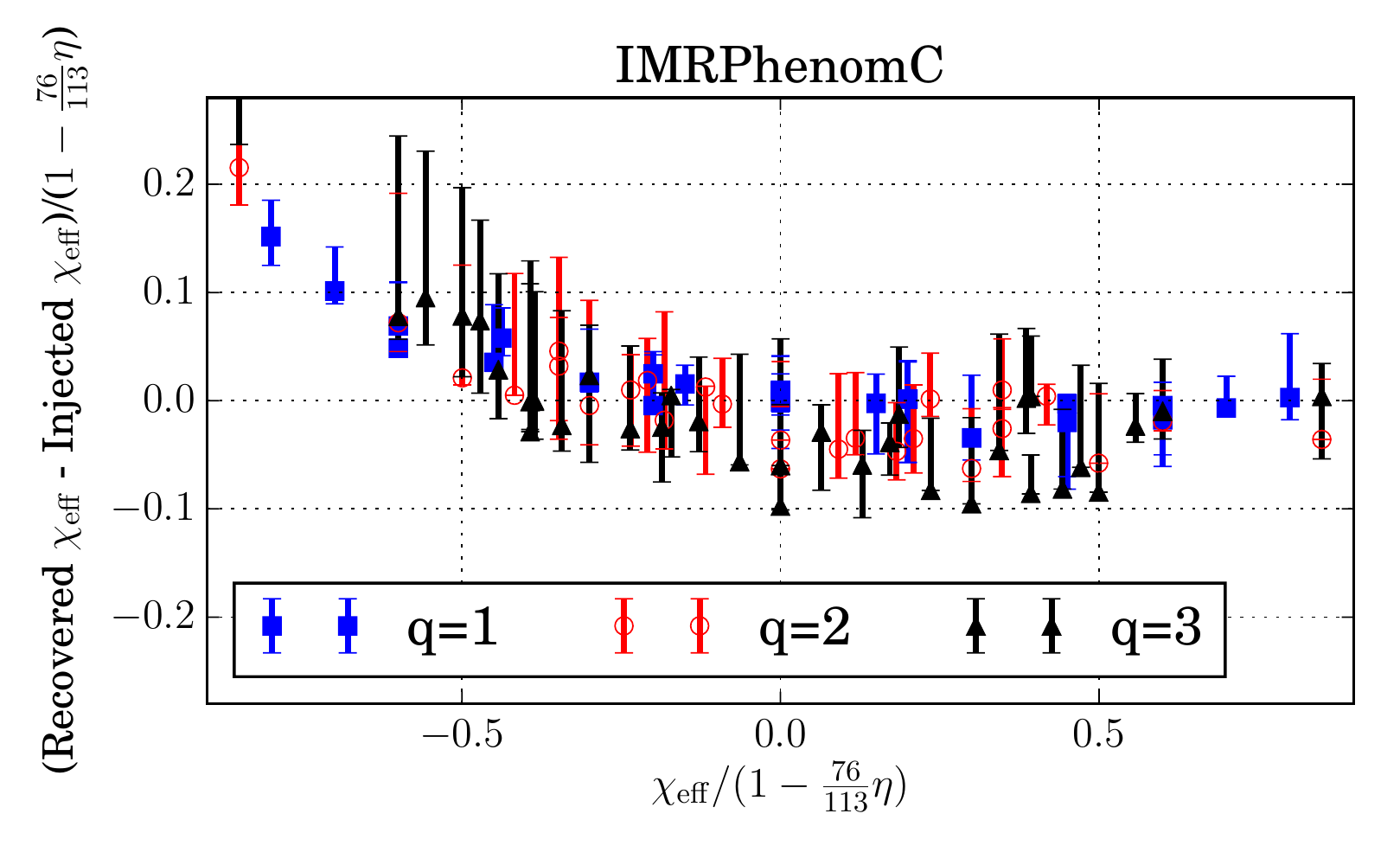}
\includegraphics[width=\columnwidth]{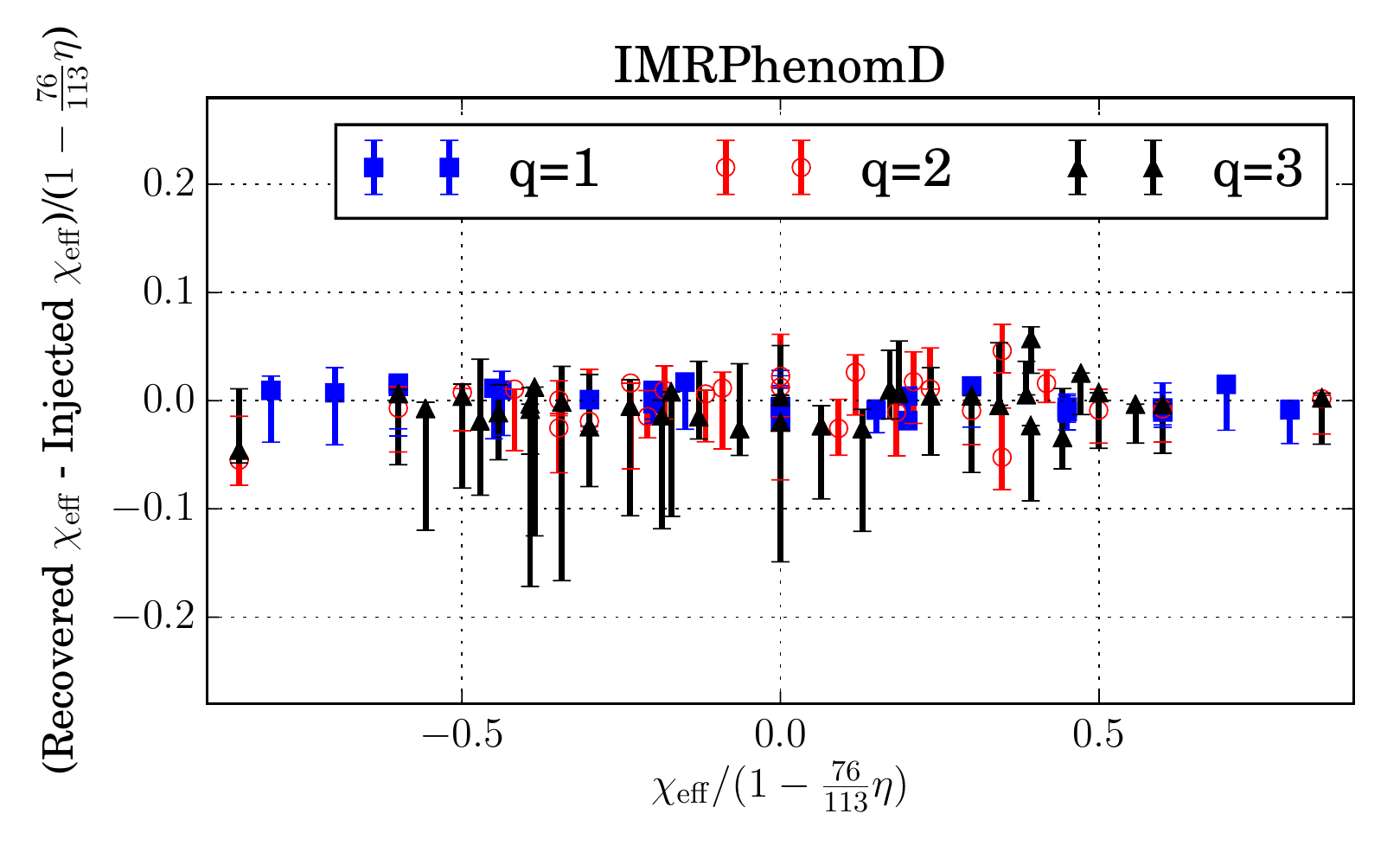}
\caption{Systematic bias in the recovery of the effective spin parameter
$\chi_\mathrm{eff}$, as a function of the normalized effective spin of the NR
waveforms. The plot-markers
show the recovered $\chi_\mathrm{eff}$ for a binary with total mass fixed at $80M_\odot$, while
the ``error-bars'' show the range spanned by the recovered $q$ as the 
injected binary mass is varied between its lowest allowed value and 
$150M_\odot$.
}
\label{fig:IMR_ChiEffError_vs_ChiEff}
\end{figure*}


Bayesian parameter estimation of BH masses and spins uses (semi-)analytical waveform
models. Its efficacy, therefore, depends critically on the 
accuracy of the waveform model used~\cite{Aasi:2013kqa}. Modeling inaccuracies
introduce systematic biases in the inferred parameter values. In this section,
we quantify these systematic parameter biases for the four waveform models considered
in this work.
To avoid the complete Markov-Chain Monte-Carlo (MCMC) procedure, we shall 
approximate the parameter bias of a waveform model as the difference in 
parameters between those parameters that maximize overlap with an NR waveform,
and the parameters of the NR waveform. We repeat this calculation for every NR
waveform. The broad features of the resulting parameter bias data are
dependent most strongly on the effective spin parameter $\chi_\mathrm{eff}$, 
and therefore, we will present results as a function of it. 
Because we project two spins $\chi_1, \chi_2$ onto the one effective spin, the
plotted data will not be single-valued. Configurations with different
$\chi_1,\chi_2$, but the same $\chi_\mathrm{eff}$ yield in general different
biases, which are plotted at their $\chi_\mathrm{eff}$ values.

First, in the left column of Fig.~\ref{fig:IMR_ChirpMassError_vs_ChiEff} we
show the fractional systematic bias in the recovery of binary chirp mass 
$\mathcal{M}_c = \frac{(m_1m_2)^{3/5}}{(m_1+m_2)^{1/5}}$ that is
intrinsic to different waveform models, as a function of the effective spin 
$\chi_\mathrm{eff}$ of the NR waveforms. In the right column of the same figure
we show the fractional biases in the recovery of binary total mass $M$.
In the top row, we show results for SEOBNRv1. The 
magnitude of the systematic biases for this model increases rapidly with 
(i) increasing magnitude of $\chi_\mathrm{eff}$, and 
(ii) increasing mass-ratio. For e.g., we see that
the recovered chirp mass can be biased by up to $15\%$ when the effective
spin is anti-aligned, while the total mass bias does not exceed $5\%$.
On the other hand, the increasing trend of systematic biases at high $\chi_\mathrm{eff}$
is to be expected since SEOBNRv1 does not support spins 
$\chi_{1,2}\geq +0.6$~\cite{Taracchini:2012}.
In the third row, we show the intrinsic bias of IMRPhenomC in recovering
binary's chirp and total masses. Focusing at the plot-markers in both panels, we 
observe that the systematic biases stay below $\sim 3\%$ for binaries with
masses at
the lower end of the mass-range probed here. However at higher masses, as with
SEOBNRv1, both the recovered chirp mass and total mass can be shifted by
$15\%$ if the binary's $\chieff \leq 0$. Relatively, the
total mass is recovered better by this model.
In comparison with SEOBNRv1, IMRPhenomC allows for less
accurate parameter recovery. 
Next, we consider the more recent SEOBNRv2 model (second row). This 
waveform model is of interest, in part, because its reduced-order 
model~\cite{Purrer:2015tud} is being used in
BBH searches being run for the presently ongoing aLIGO observing run ``O1''.
Focusing on the plot-markers we find that the systematic biases in $\mchirp$ 
recovery stay below $\sim 1-2\%$ of the true $\mchirp$ value, 
for binaries with masses $\lesssim 80M_\odot$. For higher masses ($100-150M_\odot$),
biases go up to $5\%$, but is still smaller than the statistical
uncertainty in $\mchirp$ measurement at high masses~\cite{Veitch:2015ela,Haster:2015cnn}.
The bias in $M$ has the same sign as the $\chieff$ of the binary.
Finally, in the bottom right panel, we show the results for the most recently
published IMRPhenomD model. Performing better than SEOBNRv2, IMRPhenomD furnishes
biases in the recovery of $\mchirp$ which rarely exceed $2\%$. For 
$\chieff\in[-0.6,+0.6]$ the total mass recovery does rise to $5\%$, which is
worse than the model's $\mchirp$ bias for the same signals.
We also highlight the aligned-/aligned- spin corner, where SEOBNRv2's 
mass-recovery biases rise up to $5-10\%$, while they stay 
within $2-5\%$ for IMRPhenomD. This is 
to be expected given the disagreement between SEOBNRv2 and NR in the same
region of parameter space, as shown in Sec.~\ref{s1:faithfulness}.
For all models, as illustrated in
Fig.~\ref{fig:IMR_ChirpMassError_TotalMass_Spin1z_Spin2z},
we note that the highest parameter biases for chirp mass 
correspond to the upper edge of the total-mass range probed
here, i.e. the edge of the 'error-bars' correspond to $M\sim 150M_\odot$.
In summary, for $M\leq 100M_\odot$, both SEOBNRv2 and IMRPhenomD are likely
to yield similarly accurate estimates of chirp-mass, while for higher
masses we find IMRPhenomD to be relatively more suited to parameter
estimation studies.

Further, Figure~\ref{fig:IMR_EtaRec_vs_ChiEff} shows the recovered value
of binary mass-ratio $q=m_1/m_2$, for different waveform models, as a function
of the mass-ratio and effective spin $\chi_\mathrm{eff}$ of the NR 
waveforms. As before, the plot-markers correspond to a fixed total-mass
$M=80M_\odot$, while the ``error-bars'' show the entire range of y-values
for the mass-range that we probe here (i.e. $M\in[M_\mathrm{min}, 150M_\odot]$).
In the spin-range supported by SEOBNRv1, we find that it
exhibits up to $15\%$ systematic bias in the recovery 
of $q$, with biases increasing as $|\chieff|\rightarrow 1$, i.e. for 
highly spinning binaries, including at the lower-end of the mass-range probed here. 
SEOBNRv2, on the other hand, shows a systematic trend with $\chi_\mathrm{eff}$. We 
find that the difference between the recovered and true mass-ratios increases with
$\chieff$. For negative $\chieff$, $q$ tends to be underestimated, whereas for 
positive $\chieff$, $q$ tends to be overestimated. At the high-aligned-spin end,
the mass-ratio can be over-estimated by more than $15\%$ by SEOBNRv2.
Turning to IMRPhenomC, we find that its associated $q$ bias stays within $20\%$ at 
the lower mass end, and is much larger for high binary masses. This is
particularly true for large anti-aligned $\chieff$.
IMRPhenomD, on the other hand, shows little dependency of its intrinsic
mass-ratio bias on effective spin, except that it gives slightly elevated 
$q$-bias close to $\chi_\mathrm{eff}=0$, i.e. for mixed-aligned binaries. 
Overall, we note that all models recover $q$ worse as the total-mass
of the system increases. IMRPhenomD gives relatively a better estimate
of the mass-ratio than the other models considered here.

Finally, in Fig.~\ref{fig:IMR_ChiEffError_vs_ChiEff} we show the 
bias in the recovery of the effective-spin combination, $\chi_\mathrm{eff}$,
as a function of the $\chi_\mathrm{eff}$ of the NR waveforms. 
$\chi_\mathrm{eff}$ is the leading order spin combination that enters
the binary's inspiral phasing, and therefore the matched-filter is 
expected to be most sensitive to this combination of the component
spins~\cite{Ajith:2011ec}.
Overall, we find that $\chi_\mathrm{eff}$ is well constrained,
within $\pm 0.2$ of its true value, by all the waveform models
considered. From the left column, we can compare the spin recovery of the
two older models, SEOBNRv1 and IMRPhenomC. Both of these models exhibit
strong dependence of the accuracy of spin recovery on $\chi_\mathrm{eff}$.
For SEOBNRv1, we find that its associated $\chi_\mathrm{eff}$ bias
is constrained within $\pm 0.1$ of the true value, when the source
binary's $\chi_\mathrm{eff}\leq +0.4$. When the binary's 
$\chi_\mathrm{eff}$ exceeds $+0.4$, the model gives rapidly increasing
systematic biases in its spin recovery, with $\chi_\mathrm{eff}$ being
underestimated by up to $0.25$. This trend arises because SEOBNRv1 is
restricted to component spins $\chi_{1,2}\leq +0.6$, so higher NR spin
must -- by construction -- be recovered by $\chi_{1,2}$ within SEOBNRv1's
range. IMRPhenomC exhibits a similar trend at the negative side of the 
spin range: it recovers $\chi_\mathrm{eff}$
within $\pm 0.1$ when the source's $\chi_\mathrm{eff} >-0.5$, with the
bias increasing sharply for more anti-aligned spins. 
In the top-right panel of Fig.~\ref{fig:IMR_ChiEffError_vs_ChiEff}, we
show the spin recovery by the SEOBNRv2 model. Primarily, we note that
SEOBNRv2 recovers $\chieff$ very well, with a systematic bias that stays
below $\pm 0.1$ in dimensionless spin magnitude (with rare excursions
up to $\pm 0.2$ for large aligned spins).
In addition, we note a (minor, but) interesting pattern: the bias in
spin recovery increases almost linearly with $\chieff$ between 
$-0.5\leq\chieff\leq+0.6$, 
going from $-0.1$ for $\chieff=-0.5$ to $+0.1$ for $\chieff=+0.6$.
Finally, the bottom-right panel of Fig.~\ref{fig:IMR_ChiEffError_vs_ChiEff}
shows the $\chieff$ bias for IMRPhenomD.
We find the systematic bias in $\chi_\mathrm{eff}$ associated with this 
waveform model stays between $\pm 0.1$ (as for SEOBNRv2), with best recovery
for aligned-spins and low total-masses. We also note that this bias shows
little dependence on $\chi_\mathrm{eff}$ itself, however it does increase
systematically with mass-ratio $q$ for the higher binary masses.
Overall, we find that of all the waveform models considered, both SEOBNRv2
and IMRPhenomD recover $\chi_\mathrm{eff}$ within $\pm 0.1$, with 
IMRPhenomD performing markedly more consistent for binaries with large
aligned spins.

For all models, we note that the highest parameter biases for mass and 
spin parameters correspond to the upper edge of the total-mass range probed, 
i.e. the edge of the 'errorbars' correspond to $M\sim 150M_\odot$.
We present detailed results showing the dependence of systematic parameter
biases on signal parameters in 
Appendices~\ref{as1:bias_masses},\ref{as1:bias_spins}.

From the results presented in this and previous section, we find that both
IMRPhenomD and SEOBNRv2 outperform their earlier 
incarnations in the recovery of various mass- and spin-combinations probed
here, with IMRPhenomD performing systematically better (i) at recovering
binary's chirp mass, and (ii) for parameter recovery, in general, for
systems with high aligned-spins. 

A more detailed MCMC analysis is necessary to measure the statistical
uncertainties in parameter recovery from different models in order to determine
the GW SNRs at which modeling inaccuracies will \textit{actually} begin to
dominate. Fig.~\ref{fig:SEOB_snrEff_TotalMass_Spin1z_Spin2z} only gives a lower
limit on this SNR, and we may well find that statistical uncertainties
remain dominant for even louder signals. We do, however, recommend
based on this study that
aLIGO parameter estimation efforts use either of the two waveform models
to model filters.

\section{Conclusions}\label{s1:conclusions}

LIGO and other ground-based gravitational-wave detectors rely on waveform
models for detection of compact object binaries as well as for parameter
estimation of the candidate events. Accurate waveform models are therefore
necessary to ensure high detection efficiency and to avoid systematic biases in
parameter estimation. 

Past studies focused at evaluating the accuracy of waveform models have
either used model precision as a proxy for accuracy (i.e. used model/model
discrepancy as proxy for model/true-signal discrepancy)~\cite{Damour:2000zb,
Damour2001,Damour:2002kr,Damour02,Gopakumar:2007vh,Buonanno:2009},
or have used NR simulations with zero/low-to-moderate component spins as 
benchmarks~\cite{Boyle2007,Boyle2008b,Boyle:2008,Pan2007,Hannam2007c,
Hannam:2010ec,HannamEtAl:2010,MacDonald:2012mp,Littenberg:2012uj,
Purrer:2013xma,Hinder:2013oqa,Kumar:2015tha}.
In this paper we investigate the accuracy of four
inspiral-merger-ringdown waveform models for binary black holes. Our analysis
improved in several ways over earlier work: First, we compare with numerical
relativity waveforms, rather than using the difference between analytical 
models as a proxy for their error~\cite{Damour:2000zb,
Damour2001,Damour:2002kr,Damour02,Gopakumar:2007vh,Buonanno:2009}. 
Second, the NR waveforms are independent of the investigated waveform models,
in the sense that \textit{none} of them were used in calibrating these waveform
models. Furthermore, a companion paper~\cite{Chu:2015kft} establishes
the accuracy of the NR waveforms. 
Third, we consider two recently published models, 
IMRPhenomD~\cite{Khan:2015jqa} and SEOBNRv2~\cite{Taracchini:2013rva},
the accuracy of which have not been investigated independently
(except for neutron-star black-hole binaries~\cite{Kumar:2015tha}).
Finally, our set of reference waveforms
comprehensively samples the component-spin parameter space up to
$\chi_1,\chi_2=0.9$ for $q=m_1/m_2=1$ and $0.85$ for $q=\{2,3\}$, extending
the spin coverage beyond the spins used in calibrating the waveform models.

First, we investigate the modeling accuracy of different waveform
models by computing their overlaps against our NR reference
waveforms. We rescale the NR waveforms to a range of total mass
values, from the lowest permissible (and still ensuring that it starts
at $15$~Hz, see Fig.~\ref{fig:Min_TotalMass_Spin1z_Spin2z}) up to
$m_1+m_2=150M_\odot$. From
Fig.~\ref{fig:SEOB_unfaith_TotalMass_Spin1z_Spin2z}, we find that (i)
SEOBNRv1 has overlaps above $99\%$ against NR waveforms for binaries
where the more massive black hole has spin $\chi_1 < 0.5$, which drop
to $80\%$ for larger $\chi_1$, and SEOBNRv2 performs better with
overlaps above $98\%$ across the parameter space except when both
$\chi_{1,2}$ are large and aligned. From
Fig.~\ref{fig:PhenomCD_unfaith_TotalMass_Spin1z_Spin2z}, we find (iii)
IMRPhenomC is faithful \textit{only} to NR for very mildly spinning
binaries, with overlaps falling below $90\%$ when $|\chi_2|\geq+0.3$,
and (iv) IMRPhenomD is superior to other waveform models with overlaps
(against our reference NR waveforms) above $99\%$ over the entire spin
and mass parameter space considered.  For the two most faithful models
(SEOBNRv2 and IMRPhenomD), we evaluate the indistinguishability
criterion, to find the SNR below which modeling errors do \textit{not}
significantly bias parameter estimation. From
Fig.~\ref{fig:SEOB_snrEff_TotalMass_Spin1z_Spin2z}, we find that,
except for binaries with large aligned spins on \textit{at least one
  BH}, SEOBNRv2 remains indistinguishable from real GW signals with
SNRs up to $15$ or higher.  IMRPhenomD will be indistinguishable from
real GW signals with SNRs of $30$ and above for equal-mass, equal-spin
binaries, and for SNRs $\gtrsim 15$ over most of the remaining
parameter space.  These SNR ranges are very likely to be conservative,
due to the overly strict nature of the distinguishability criterion
used~\cite{Lindblom2008}.

Second, we investigate the effectualness of different waveform models (including
two additional PN-based ones) for use as aLIGO BBH detection filters. 
Detection searches have an additional degree of freedom: the recovered SNR is
maximized over the mass and spin parameters that characterize model 
waveforms. We compute the fitting factors~\cite{FittingFactorApostolatos} of 
different waveform models against our NR waveforms, to measure the SNR loss
due to modeling inaccuracies in isolation. As shown in 
Fig.~\ref{fig:combined_mismatches_TotalMass_Spin1z_Spin2z_IMR}, we find that
(i) SEOBNRv1 is effectual over the entire parameter range it supports, i.e. for
$\chi_{1,2}\leq +0.6$, with fitting factors higher than $99.5\%$; (ii) SEOBNRv2;
has fitting factors above $99.5\%$ across the considered region of the parameter
space, except for the 
high-spin/high-spin corner, where its fitting factors fall to 97\%;
(iii) IMRPhenomC recovers $99+\%$ of the SNR over most of the parameter space, 
except when both holes have either large aligned or large anti-aligned spins, 
in which cases it still recovers $98+\%$ of the optimal SNR; and (iv) 
IMRPhenomD out-performs all other waveform models with fitting factors
above $99.5\%$ over the \textit{entire} parameter range probed. 
We note that the
frequency domain IMRPhenomC model makes good use of the intrinsic degeneracy
in the waveform parameter space, and is therefore well suited to detection
searches. SEOBNRv2, on the other hand, does not compensate for its inaccuracy
in the high-spin/high-spin corner of the parameter space with modified 
intrinsic parameters, and will likely need to be re-calibrated there.

Third, we investigate the systematic biases in parameter recovery caused by
intrinsic model inaccuracies. We find that 
(i) both IMRPhenomD and SEOBNRv2 recover
binary \textit{chirp-mass} to within $\pm 2\%$ for $M\gtrsim 70M_\odot$,
and $\pm 5-7\%$ for $M\gtrsim 110M_\odot$, with IMRPhenomD systematically
more accurate for aligned spins. 
(ii) Binary total mass is recovered with somewhat larger systematic
biases across the mass range, spanning $\pm 5\%$ for binaries for which the 
chirp mass is recovered within $\pm 2\%$.
(iii) SEOBNRv2 and IMRPhenomD recover the binary \textit{mass-ratio} with 
comparable accuracy (within $\pm 10-15\%$), with IMRPhenomD showing smallest
biases for aligned spin binaries.
Finally, (iv) the leading order PN spin combination $\chieff$ is the best
recovered with IMRPhenomD (within $\pm 0.1$), followed closely by SEOBNRv2.
The remaining two models show larger biases for all intrinsic parameters
(see Fig.~\ref{fig:IMR_ChiEffPNError_TotalMass_Spin1z_Spin2z})

In summary, we find that the more recently published SEOBNRv2 and
IMRPhenomD models reproduce NR waveforms with identical parameters
more accurately than their earlier counterparts, and have very good
SNR recovery.  We also find that the frequency-domain IMRPhenomC model
is effectual enough for aLIGO detection searches aimed at
comparable-mass aligned spin high-mass BBHs. We recommend that aLIGO
parameter estimation efforts prefer IMRPhenomD or SEOBNRv2 as the
waveform model of choice, in favor other currently available frequency
and time domain waveform models.

As noted previously, the parameter biases estimated here need to be 
comprehensively compared
with the statistical errors in parameter recovery from detailed MCMC 
analyses, in order to determine the actual GW SNRs where modeling errors
begin to dominate over other error sources of uncertainty. A recent
study~\cite{Purrer:2015nkh} indicates that such might be the case for 
SNRs $\approx 20-30$ and higher (for SEOBNRv2).
We also note
that in order to thoroughly sample the spin parameter space, we have
restricted ourselves to small mass-ratios, i.e. $q=\{1,2,3\}$. The results
presented here are therefore applicable to comparable-mass BBHs with total
masses $M\gtrsim 50M_\odot$, and will
be extended to higher mass-ratios and lower total-masses in the future,
as longer and higher $q$ simulations become less computationally 
expensive with advances in NR technology~\cite{Szilagyi:2015rwa}.
Finally, we use the dominant quadrupolar multipoles here of the reference
NR waveforms, and leave a study of the sub-dominant modes for future
work. We expect their effect to be limited to the highest masses and 
mass-ratios considered here~\cite{Graff:2015bba}, although a more rigorous
treatment is needed to re-affirm this conclusion.

The relative performance of the four models considered very
  closely matches their relative age, with the newest model (PhenomD)
  performing best.  This is expected, given the rapid progress in
  numerical relativity and waveform modeling.  We also expect the
  trend towards higher accuracy to continue with future models (e.g.,
  an EOB model calibrated to new NR simulations).  However, our tests
  show that for the comparatively simple part of parameter space
  considered here, recent models are already very good.  Even the
  older models considered here (SEOBNRv1 and PhenomC) are remarkably
  good, with very small loss of SNR for event detection
  (cf. Fig.~\ref{fig:combined_mismatches_TotalMass_Spin1z_Spin2z_IMR}).
  This is a remarkable success, given the sparse parameter space
  coverage, cf. Fig.~\ref{fig:NRParams_IMR}.  Future work should be
  directed toward expanding parameter space coverage (higher
  mass-ratios, precessing spins), and to include subdominant modes.

\begin{acknowledgments}
We thank Kipp Cannon, Adam Lewis, Eric Poisson and Aaron
    Zimmerman for helpful discussions.  We are grateful to Ofek
    Birnholtz, Sebastian Khan, Lionel London, Frank Ohme and Michael
    P{\"u}rrer, for providing access to the IMRPhenomD code. 
  Simulations used in this work were performed with
  \texttt{SpEC}~\cite{SpECwebsite}.  We gratefully acknowledge support
  for this research at CITA from NSERC of Canada, the Ontario Early Researcher Awards Program, the Canada Research
  Chairs Program, and the Canadian Institute for Advanced Research; at
  Caltech from the Sherman Fairchild Foundation and NSF grants
  PHY-1404569 and AST-1333520; at Cornell from the
  Sherman Fairchild Foundation and NSF grants PHY-1306125 and
  AST-1333129; and at Princeton from NSF grant PHY-1305682 and the
  Simons Foundation.  Calculations were performed at the GPC
  supercomputer at the SciNet HPC Consortium~\cite{scinet}; SciNet is
  funded by: the Canada Foundation for Innovation (CFI) under the
  auspices of Compute Canada; the Government of Ontario; Ontario
  Research Fund (ORF) -- Research Excellence; and the University of
  Toronto. Further calculations were performed on the Briar\'ee
  cluster at Sherbrooke University, managed by Calcul Qu\'ebec and
  Compute Canada and with operation funded by the Canada Foundation
  for Innovation (CFI), Minist\'ere de l'\'Economie, de l'Innovation
  et des Exportations du Quebec (MEIE), RMGA and the Fonds de
  recherche du Qu\'ebec - Nature et Technologies (FRQ-NT); on the
  Zwicky cluster at Caltech, which is supported by the Sherman
  Fairchild Foundation and by NSF award PHY-0960291; on the NSF XSEDE
  network under grant TG-PHY990007N; on the NSF/NCSA Blue Waters at
  the University of Illinois with allocation jr6 under NSF PRAC Award
  ACI-1440083.  H.P. and P.K. thank the Albert-Einstein Institute,
  Potsdam, for hospitality during part of the time where this research
  was completed.

\end{acknowledgments}

\begin{appendix}\label{as:appendix}

\section{Bias in mass combinations}\label{as1:bias_masses}
\begin{figure*}
\centering
    
\includegraphics[width=1.8\columnwidth]{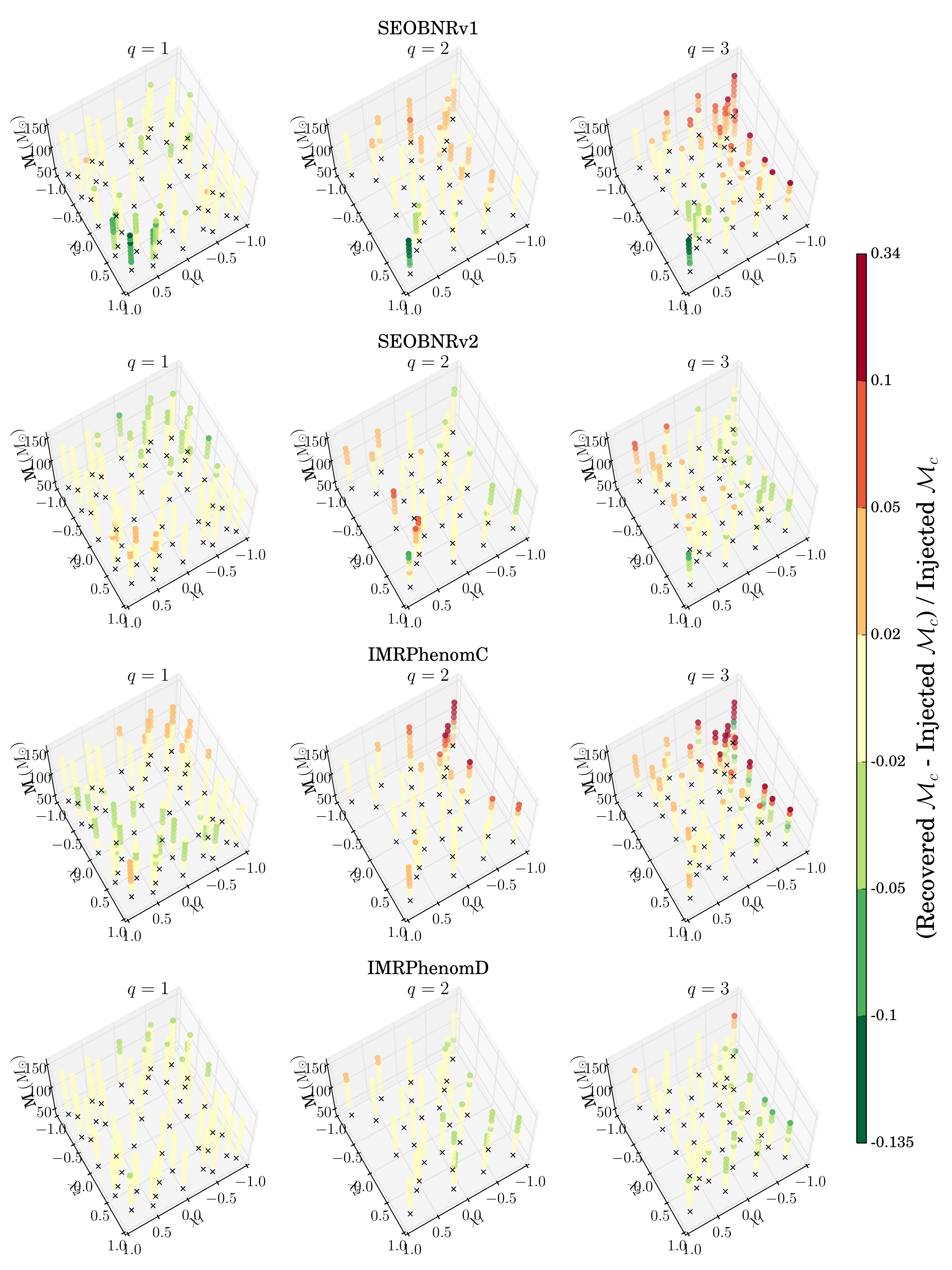}\\
\caption{Systematic bias in the recovered chirp mass 
$\mathcal{M}_c$ for each waveform model, compare to
Fig.~\ref{fig:IMR_ChirpMassError_vs_ChiEff}. As in Fig.~\ref{fig:combined_mismatches_TotalMass_Spin1z_Spin2z_IMR}, 
the black crosses denote the values of component spins in the $x-y$ plane.
Biases below $2\%$ are shown nearly transparently, to emphasize regions 
with larger biases.
}
\label{fig:IMR_ChirpMassError_TotalMass_Spin1z_Spin2z}
\end{figure*}

\begin{figure*}
\centering
    
\includegraphics[width=1.8\columnwidth]{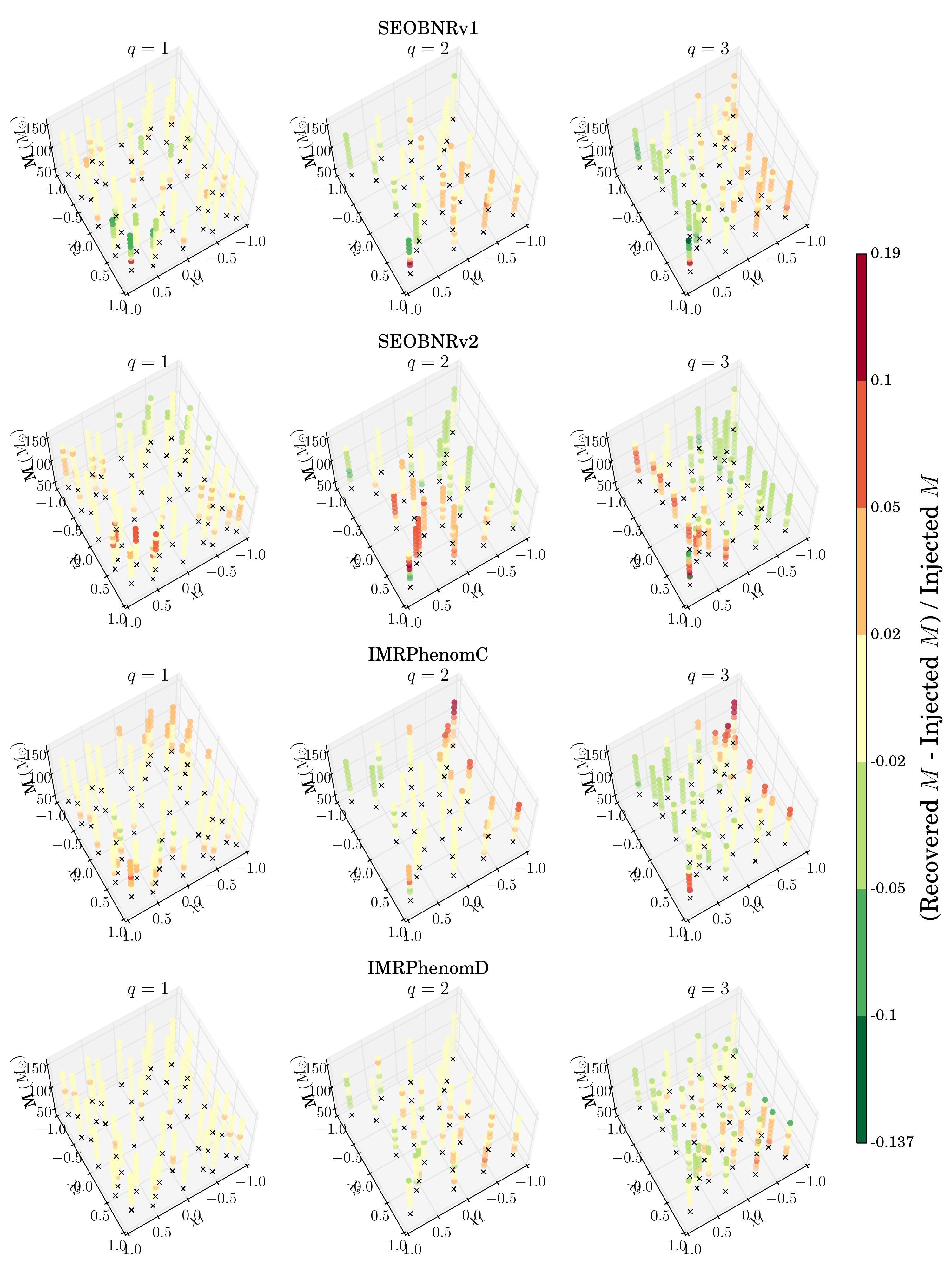}\\
\caption{Systematic bias in the recovered total mass 
$M$ for each waveform model, compare to
Fig.~\ref{fig:IMR_ChirpMassError_vs_ChiEff}. As in Fig.~\ref{fig:combined_mismatches_TotalMass_Spin1z_Spin2z_IMR}, 
the black crosses denote the values of component spins in the $x-y$ plane.
Biases below $2\%$ are shown nearly transparently, to emphasize regions 
with larger biases.
}
\label{fig:IMR_TotalMassError_TotalMass_Spin1z_Spin2z}
\end{figure*}

\begin{figure*}
\centering
    
\includegraphics[width=1.8\columnwidth]{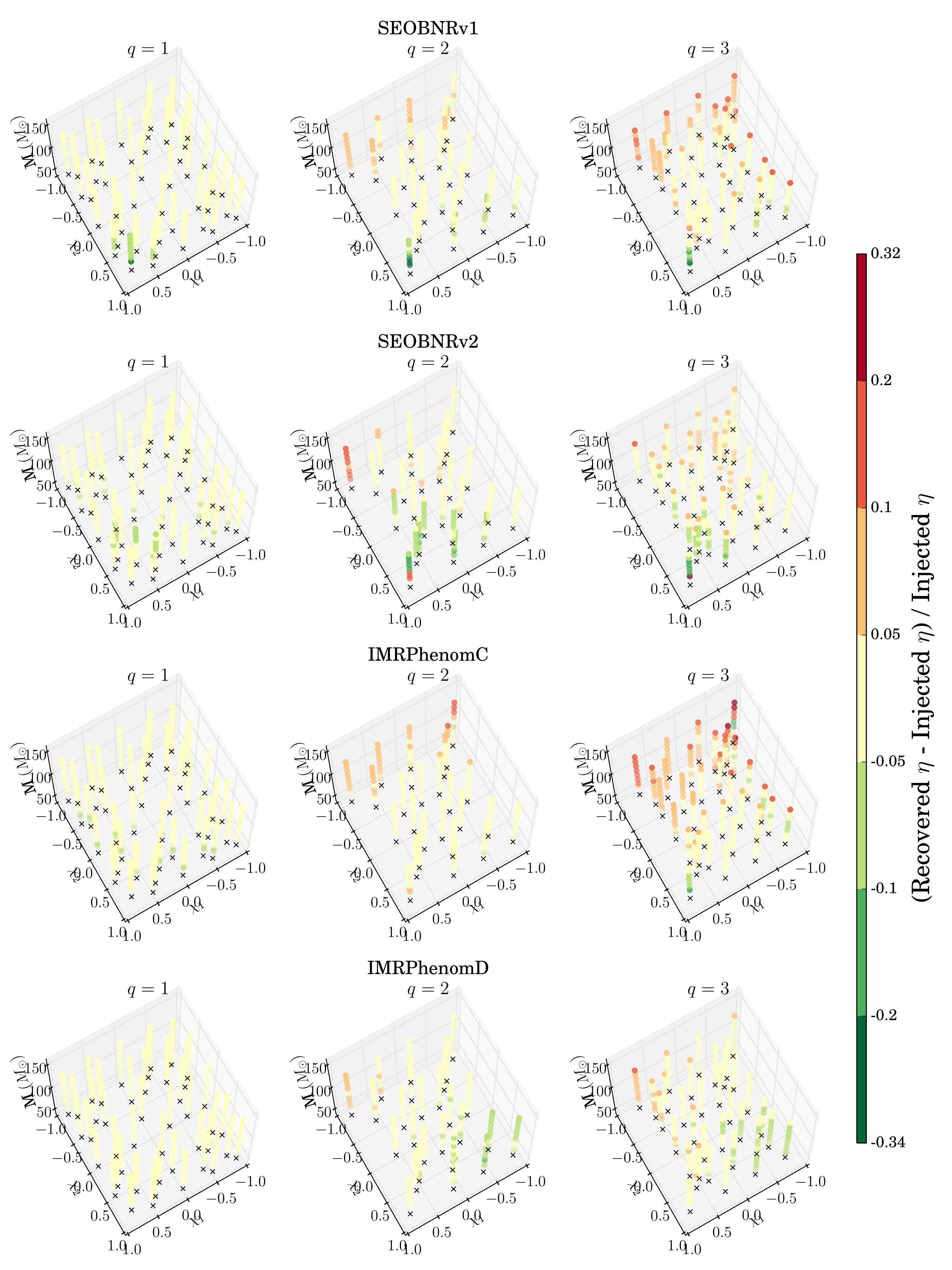}
\caption{Systematic bias in the recovered symmetric mass-ratio $\eta$
for the considered waveform models. As in Fig.~\ref{fig:combined_mismatches_TotalMass_Spin1z_Spin2z_IMR}, 
the black crosses denote the values of component spins in the $x-y$ plane.
}
\label{fig:IMR_EtaError_TotalMass_Spin1z_Spin2z}
\end{figure*}

In this appendix, we present additional information about the parameter
estimation mass recovery.
Fig.~\ref{fig:IMR_ChirpMassError_TotalMass_Spin1z_Spin2z} shows the 
chirp-mass recovery as a function of both component spins, expanding on 
the left column of
Fig.~\ref{fig:IMR_ChirpMassError_vs_ChiEff} in the main text.
Fig.~\ref{fig:IMR_TotalMassError_TotalMass_Spin1z_Spin2z} shows the 
total-mass recovery, similarly expanding on the right column of
Fig.~\ref{fig:IMR_ChirpMassError_vs_ChiEff}.
Fig.~\ref{fig:IMR_EtaError_TotalMass_Spin1z_Spin2z} plots the recovery of
symmetric mass-ratio $\eta$ (c.f. Fig.~\ref{fig:IMR_EtaRec_vs_ChiEff}).

From Fig.~\ref{fig:IMR_ChirpMassError_TotalMass_Spin1z_Spin2z}, we find 
that (i) for the \textit{least} massive binaries considered, both SEOBNRv2 and
IMRPhenomD introduce less than $2\%$ systematic biases in the recovery
of binary chirp mass; with the same rising to $10\%$ for the \textit{most}
massive binaries. (ii) The chirp-mass bias measured for SEOBNRv1 closely
follows that of SEOBNRv2, except when both black holes carry large spins
(both aligned and anti-aligned) -- where the bias exceeds $10\%$.
(iii) IMRPhenomC, on the other hand, has intrinsic chirp-mass biases
that remain below $5\%$ over the considered parameter space, except when
the more massive hole has large anti-aligned spin -- for which the biases
exceed $10\%$ for binary mass $M\gtrsim 100M_\odot$.
From Fig.~\ref{fig:IMR_TotalMassError_TotalMass_Spin1z_Spin2z}, we find that
(i) both SEOBNRv1 and IMRPhenomC incur smaller systematic biases in $M$
recovery than they do for $\mchirp$ recovery, especially for large 
anti-aligned and aligned spins.
(ii) SEOBNRv2 shows the opposite pattern, i.e. it recovers $M$ with \textit{more}
accuracy than $\mchirp$, especially for larger mass-ratios and larger
spins on the bigger black hole. Finally, (iii) IMRPhenomD recovers both mass 
combinations with relatively the highest accuracy.

Further onto $\eta$ recovery, the first thing we note from
Fig.~\ref{fig:IMR_EtaError_TotalMass_Spin1z_Spin2z}, is that all four models
recover $\eta$ well (within $2\%$) for equal-mass binaries, and this fidelity
decreases as we go from $q=1\rightarrow 3$. The only exception is SEOBNRv1 at
spins outside the range of the model (i.e. $\chi_{1,2}>+0.6$).
For $q=2$, we find that (i) the biases intrinsic to SEOBNRv2 are higher than 
SEOBNRv1, reaching $15-20\%$ and $10-15\%$, respectively, for both. SEOBNRv2
also gives a 
systematic under-estimation of $\eta$ by $-15\%$ when both holes have large
positive-aligned spins. (ii) IMRPhenomC, in contrast, performs better with 
biases staying below $10\%$, even at the highest binary masses. And, (iii)
IMRPhenomD shows the highest fidelity (with $\eta$ biases below $5\%$).
Increasing the mass-ratio to $q=3$, we find that (i) all three models other than
IMRPhenomD manifest larger than $10\%$ systematic biases in $\eta$ recovery.
(ii) For SEOBNRv1 the $\eta$ bias increases as the spin on the smaller hole
becomes increasingly anti-aligned, while SEOBNRv2 over-estimates $\eta$ by 
$5-10\%$ for aligned BH spins (this trend was already apparent in
Fig.~\ref{fig:IMR_EtaRec_vs_ChiEff}). (iii) IMRPhenomC shows relatively the
worst $\eta$ recovery of the four with biases ranging from $-15\%$ to $20\%$.
IMRPhenomD confirms our earlier results and is found to perform best
at $\eta$ recovery, significantly improving upon its predecessor IMRPhenomC.

\section{Biases in recovered spins}\label{as1:bias_spins}
\begin{figure*}[ht]
\centering    
\includegraphics[width=1.8\columnwidth]{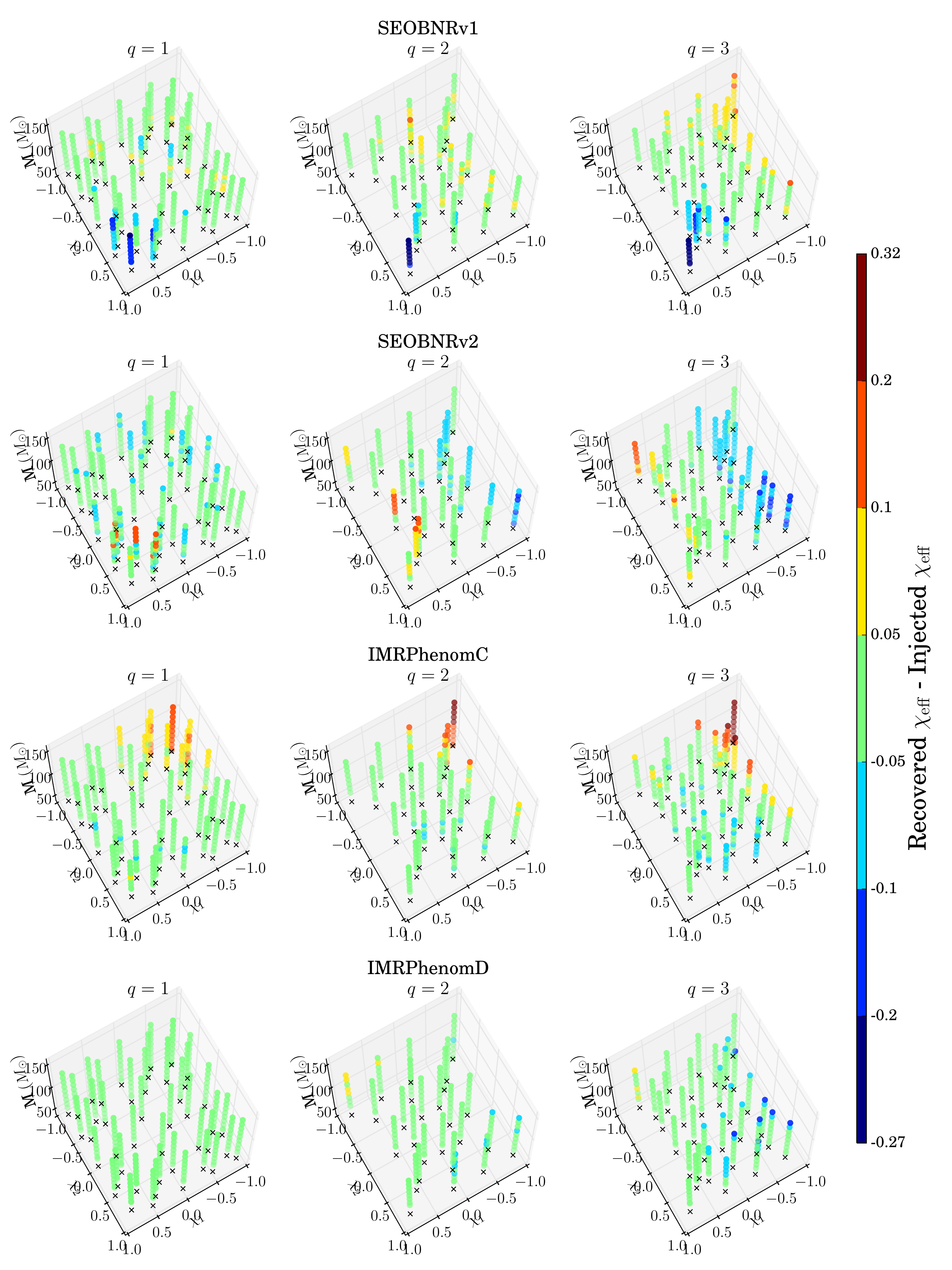}    
\caption{Systematic bias in the recovered values of the 1.5PN effective spin
$\chi_\mathrm{eff}$, for the SEOBNRv1, SEOBNRv2, IMRPhenomC, and IMRPhenomD
models (from top to bottom).
}
\label{fig:IMR_ChiEffPNError_TotalMass_Spin1z_Spin2z}
\end{figure*}
\begin{figure*}[ht]
\centering    
\includegraphics[width=1.8\columnwidth]{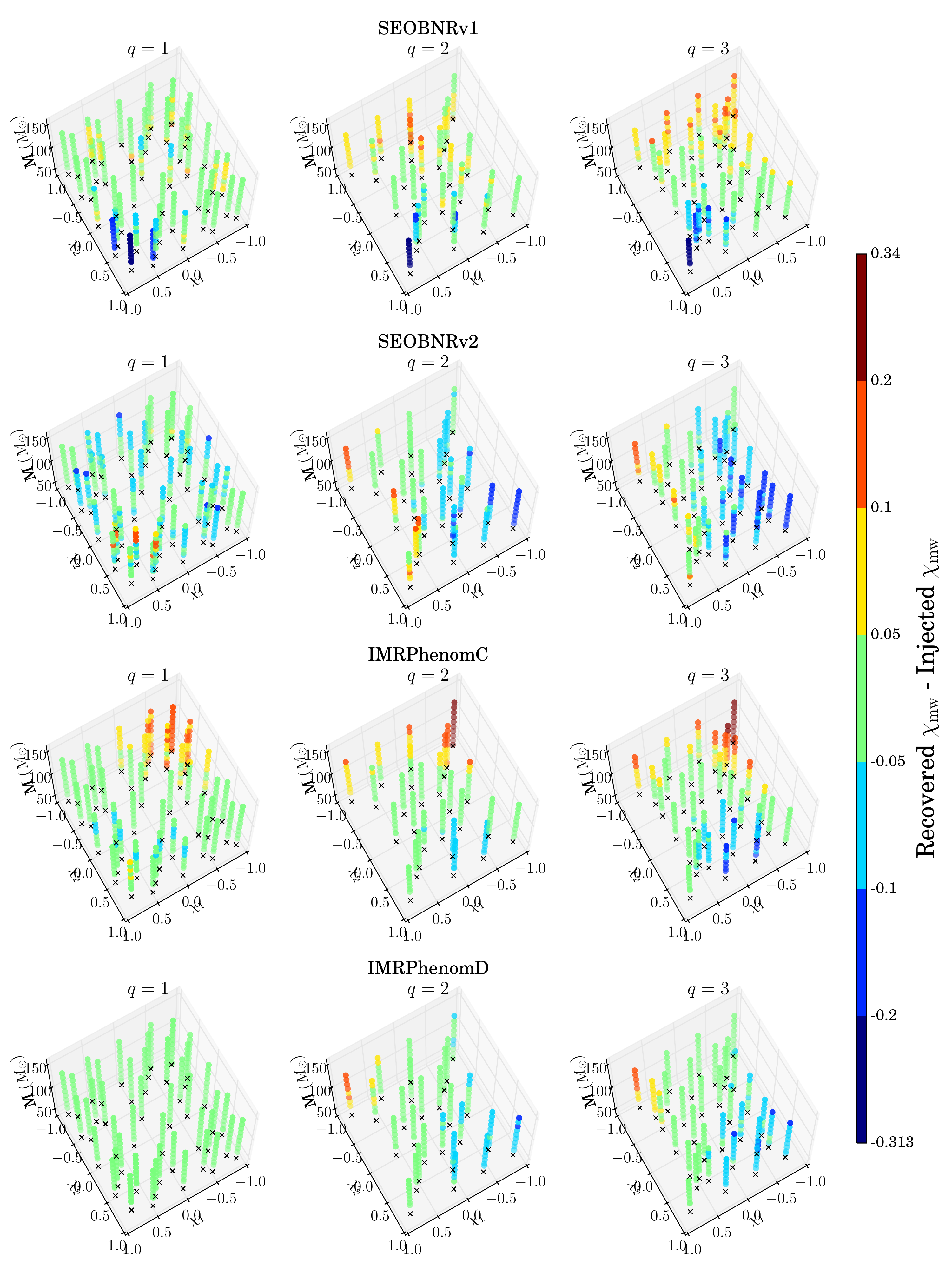}
\caption{Systematic bias in the recovered values of the mass-weighted 
effective spin $\chi_\mathrm{mw}$, for the SEOBNRv1, SEOBNRv2, IMRPhenomC,
and IMRPhenomD models (from top to bottom).
}
\label{fig:IMR_ChiMWError_TotalMass_Spin1z_Spin2z}
\end{figure*}

In this appendix, we present additional information about the parameter
estimation spin recovery.
Fig.~\ref{fig:IMR_ChiEffPNError_TotalMass_Spin1z_Spin2z} 
and~\ref{fig:IMR_ChiMWError_TotalMass_Spin1z_Spin2z} show the bias in the
recovery of two different spin parameters, the effective spin $\chieff$
(c.f. Eq.~\ref{eq:chieff}) and the mass-weighted spin $\chi_\mathrm{mw}$
(c.f. Eq.~\ref{eq:chimw}). 

The overall trends are similar for $\chieff$ and $\chi_\mathrm{mw}$ :
All four models recover $\chieff$ well, with absolute systematic biases
between $\pm 0.2$. Of the four, IMRPhenomD stands out by recovering 
$\chieff$ to within $\pm 0.1$ of the true value. SEOBNRv2
follows very closely, with $\chieff$ biases rising higher than
$0.1$ only for very massive binaries (with $M\gtrsim 100M_\odot$) with
large spins (magnitude) on at least one hole. Both of the two remaining
models show a strong correlation between the $\chieff$ bias and the 
$\chieff$ of the binary itself. While SEOBNRv1 \textit{under-}estimates 
$\chieff$ by up to $0.25$ when both holes have large \textit{aligned} spins, 
IMRPhenomC \textit{over-}estimates $\chieff$ when both holes have large
\textit{anti-aligned} spins. 
Next, we focus on $\chi_\mathrm{mw}$. As for $\chieff$, IMRPhenomD was found to
recover $\chimw$ with the smallest biases, which only exceed $\pm 0.1$ for
unequal-mass binaries with aligned (anti-aligned) spin on the larger (smaller)
hole. As this is the spin mapping used by IMRPhenomC to capture component spin
effects on phasing, we notice from
Fig.~\ref{fig:IMR_ChiEffPNError_TotalMass_Spin1z_Spin2z} that it also recovers
$\chi_\mathrm{mw}$ very well -- except when both components have large
anti-aligned spins, in which case it overestimates $\chimw$ by up to $0.3$
dimensionless units.
Of the two EOB models, SEOBNRv2 recovers $\chimw$ better with systematic
biases increasing with mass-ratio $q$, but not exceeding $\pm 0.2$.
SEOBNRv1, on the other hand, shows the inverse pattern of IMRPhenomC, giving
large systematic biases in $\chimw$ for binaries with $\chi_{1,2}\geq 0.6$ --
which is expected by construction from the model as it does not support these 
component spins.

Overall, we find both IMRPhenomD and SEOBNRv2 models viable for aLIGO parameter
estimation studies aimed at high-mass binary black holes with non-precessing
spins

\end{appendix}

\section*{References}
\bibliography{References-prd}

\end{document}